\documentclass[letterpaper,12pt]{article}

\usepackage{amssymb,latexsym,amsmath}
\usepackage{fullpage,mathrsfs}

\makeatletter \@addtoreset{equation}{section}
\makeatother

\def\be{\begin{equation}}
\def\ee{\end{equation}}
\def\bea{\begin{eqnarray}}
\def\eea{\end{eqnarray}}
\def\de{\partial}
\newcommand{\nn}{\nonumber}

\newcommand{\diff}{\mathrm{d}}

\newcommand{\pontry}{P}
\newcommand{\euler}{E}
\newcommand{\weylsq}{C^2}
\newcommand{\imf}{\mathcal{G}}
\newcommand{\ii}{\mathrm{i}} 
\newcommand{\rS}{r} 

\usepackage[pdftex]{graphicx}

\numberwithin{equation}{section}       

\begin{document}

\begin{titlepage}

\begin{center}

\today

\vskip 2.3 cm 

\vskip 5mm

{\Large \bf Supersymmetry on curved spaces and\\[3mm] 

superconformal anomalies}

\vskip 15mm

{Davide Cassani and Dario Martelli}

\vskip 1cm

\textit{Department of Mathematics, King's College London, \\ [1mm]
The Strand, London WC2R 2LS,  United Kingdom\\}

\end{center}

\vskip 2 cm

\begin{abstract}
\noindent We study the consequences of unbroken rigid supersymmetry of four-dimensional field theories placed on curved manifolds. We show that in Lorentzian signature the background vector field coupling to the R-current is determined by the Weyl tensor of the background metric. 
In Euclidean signature, the same holds if two supercharges of opposite R-charge are preserved, otherwise the (anti-)self-dual part of the vector field-strength is fixed by the Weyl tensor.  As a result of this relation, the trace and R-current anomalies of superconformal field theories simplify, with the trace anomaly becoming purely topological. In particular, in Lorentzian signature, or in the presence of two Euclidean supercharges of opposite R-charge, supersymmetry of the background implies that the term proportional to the central charge $c$ vanishes, both in the trace and R-current anomalies. This is equivalent to the vanishing of a superspace Weyl invariant. We comment on the implications of our results for holography.
\end{abstract}

\end{titlepage}

\pagestyle{plain}
\setcounter{page}{1}
\newcounter{bean}
\baselineskip18pt


\baselineskip 6 mm

\newpage


\section{Introduction}

The Weyl anomaly has a very rich history \cite{Duff:1993wm,dufftalk} and many different applications. This arises when a conformal field theory is
placed on a curved space, and is measured by certain curvature invariants contributing to the trace of the energy-momentum tensor. Four-dimensional superconformal field theories possess an R-symmetry, which also becomes anomalous on generic curved backgrounds.  
 The underlying supersymmetry of the theory relates the terms appearing in the trace and R-symmetry anomalies, which in fact belong to the same supermultiplet.
 For an ${\cal N}=1$ superconformal theory coupled to an arbitrary background metric $g_{mn}$ and gauge field $A_m$ (sourcing the R-symmetry current), 
the anomalies read~\cite{Anselmi:1997am}\footnote{Note that (\ref{correcttrace}) and  (\ref{correctchiral})  correct errors in the formulae presented in reference \cite{Anselmi:1997am}. We thank Dan Freedman for correspondence on this point.
} (see also~\cite{Osborn}) 
\bea\label{correcttrace}
 T_m^m   & = & \frac{c}{16\pi^2}  \weylsq   \,- \,
\frac{a}{16\pi^2}  \euler \, - \,  \frac{c}{6\pi^2}\,   F^2  ~, \\ [3mm]
\nabla_m  J^m  & = & \frac{c-a}{24\pi^2}\, \pontry\, + \,\frac{5a-3c}{27\pi^2} \,F \widetilde F\,,
 \label{correctchiral}
 \eea
where $c$ and $a$ are the central charges of the theory, $\weylsq$, $\euler$ and $\pontry$ denote the Weyl, Euler and Pontryagin invariants respectively 
(whose precise definition will be given later), while $F = \diff A$ and $\widetilde F$ is its Hodge dual.  In the present paper, we will show that these anomaly formulae simplify when the background preserves some supersymmetry.

In \emph{a priori} unrelated recent developments, there has been interest in coupling supersymmetric field theories 
to a curved metric and other background fields, in such a way that some suitably defined rigid supersymmetry is preserved. The main motivation for considering these 
deformations is that
supersymmetric field theories on compact manifolds are often amenable to localization techniques, which may be used to compute exactly various observables, at any value of the couplings~\cite{Pestun:2007rz}.
Starting with~\cite{FestucciaSeiberg}, this led to systematic studies of the conditions for a background to support rigid supersymmetry in different models and dimensions, both in Euclidean~\cite{SamtlebenTsimpis,KTZ,Dumitrescu:2012ha,Liu:2012bi,Dumitrescu:2012at,KehagiasRusso,Closset:2012ru,Samtleben:2012ua} and in Lorentzian~\cite{Cassani:2012ri,Liu:2012bi,deMedeiros:2012sb,Hristov:2013spa} signature (see also~\cite{Blau:2000xg} for some earlier work, and~\cite{KuzenkoSymmetries} for a superspace perspective). 
In this paper we will focus on four-dimensional spaces, either Lorentzian or Riemannian. 

Three approaches to rigid supersymmetry in four dimensions have been discussed so far in the literature, based on old minimal~\cite{FestucciaSeiberg,SamtlebenTsimpis,Liu:2012bi,Dumitrescu:2012at}, new minimal~\cite{KTZ,Dumitrescu:2012ha,Cassani:2012ri} and conformal~\cite{KTZ,Cassani:2012ri} supergravity. Although these encode slightly different notions of supersymmetry, it turns out that in all cases, and both in Euclidean~\cite{KTZ} and Lorentzian~\cite{Cassani:2012ri} signature, unbroken supersymmetry implies the existence of a solution $\zeta_+$ to the \emph{charged conformal Killing spinor} (CKS) equation:
\be\label{ChargedCKS}
\nabla_m^A \zeta_+ \ = \ \frac{1}{4}\gamma_m  \nabla^A \!\cdot  \zeta_+\, .
\ee
Here $\zeta_+$ is a chiral spinor charged under the background gauge field $A_m$ coupling to the field theory R-current, with $\nabla_m^A\zeta_+ \equiv \big(\nabla_m - \ii A_m \big)\zeta_+$. The term $\nabla^A \!\cdot \zeta_+\equiv \gamma^n \nabla_n^A\zeta_+$ makes the equation traceless and conformal invariant.
This is also known as the (charged) twistor spinor equation~\cite{PenroseRindlerVol2}. 
In references~\cite{KTZ,Dumitrescu:2012ha} it was shown that in Euclidean signature a nowhere vanishing solution to~\eqref{ChargedCKS} exists if and only if there exists a Hermitian metric, while in reference \cite{Cassani:2012ri} it was found that in Lorentzian signature a solution is equivalent to the existence of a null conformal Killing vector. 
In both cases, the gauge field $A$ can be determined explicitly from the metric data, although in a rather intricate way. 

In this paper, we will revisit the analysis of the CKS equation by studying its integrability condition. Combining the technique of $G$-structures with the Newman--Penrose formalism, we will demonstrate that in Lorentzian signature the gauge curvature can be extracted completely from the conformally invariant part of the spacetime curvature, namely from the \emph{Weyl tensor} of the metric. This also holds in Euclidean signature when two supercharges of opposite R-charge are preserved. 

Our characterization leads to precise relations among the curvature invariants which appear in the superconformal anomalies, implying that the latter simplify on supersymmetric backgrounds, with (\ref{correcttrace}) becoming purely topological, as is (\ref{correctchiral}). In particular, in Lorentzian signature, or in Euclidean backgrounds possessing two conformal Killing spinors with opposite R-charge, we obtain 
\bea
  \weylsq    -   \frac{8} {3}   F^2  \ = \  0 \ = \
\pontry - \frac{8}{3} F \widetilde F  \,,
\eea
so the contributions proportional to the central charge $c$ cancel out both in (\ref{correcttrace}) and (\ref{correctchiral}). These relations are equivalent to the vanishing of the bosonic part of a super-Weyl invariant, whose real part is also the Lagrangian of conformal supergravity~\cite{FradkinTseytlin}.

The paper is organised as follows. In section \ref{ckseq} we discuss the integrability condition of the CKS equation~(\ref{ChargedCKS}), yielding the relation between the Maxwell and Weyl background tensors. In section~\ref{mainsection} we discuss the implications on the superconformal anomalies. In section~\ref{theexample} we illustrate the results by presenting backgrounds with $SU(2)\times U(1)^2$ symmetry that are free of anomalies.
In section \ref{holosection} we review the holographic realisation of the superconformal anomalies. Section~\ref{discutiamo} concludes. Appendix~\ref{ConventionsApp} gives our conventions, Appendix~\ref{DerivationIntegrability} derives the integrability condition, while in Appendix~\ref{reviewanomalies}
we comment on the anomaly formulae~(\ref{correcttrace}) and~(\ref{correctchiral}).


\section{Integrability of the CKS equation}
\label{ckseq}

In this section we study the integrability of the CKS equation. The analysis is performed by combining the method of $G$-structures with the Newman--Penrose tetrad formalism. The result will be a set of conditions relating the different components of the Weyl and Maxwell tensors. 

Before starting our technical analysis, let us recall the different notions of rigid supersymmetry on four-dimensional curved manifolds that have been studied recently. These differ in the choice of the supercurrent multiplet being employed to couple the theory to background supergravity. 
If the flat-space field theory is superconformal, then it is naturally coupled to conformal supergravity. This is also seen from holography, as bulk gauged supergravity on an asymptotically locally AdS space induces conformal supergravity on the boundary~\cite{Balasubramanian:2000pq} (with the conformal supergravity action being mapped into the field theory trace anomaly~\cite{Liu:1998bu,Henningson:1998gx}, see section~\ref{holosection} for more details).  The condition ensuring that the field theory preserves at 
least one supercharge arises from requiring that the gravitino variation vanishes on the background, and in this case is found to be the CKS equation~\eqref{ChargedCKS} \cite{KTZ,Cassani:2012ri}.
 In Euclidean signature, the supersymmetry parameters $\zeta_+$ and $\widetilde\zeta_-$ that in Lorentzian signature would be charge conjugate to each other transform under different $SU(2)$ subgroups of $Spin(4)$ and are thus independent. For this reason, one should regard the equation
\be\label{ChargedCKSminus}
\nabla_m^A \widetilde\zeta_- \ = \ \frac{1}{4}\gamma_m \nabla^A\!\cdot  \widetilde\zeta_-\,,
\ee
which in Lorentzian signature would merely be the charge conjugate of~\eqref{ChargedCKS}, as independent. Here, $\widetilde\zeta_-$ is a spinor of negative chirality with R-charge opposite to $\zeta_+$: $\nabla^A\widetilde\zeta_-= (\nabla +\ii A)\widetilde\zeta_-$.
Moreover, in Euclidean signature the gauge field $A$ is allowed to take complex values (while we will keep the metric real).

If the field theory is supersymmetric but not necessarily superconformal, it may be coupled  to old minimal supergravity through its Ferrara--Zumino supercurrent or, if it has an R-symmetry, to new minimal supergravity via its R-multiplet. Although the requirements for a supersymmetric background are slightly different in these cases, the CKS equation turns out to always be a necessary condition. Therefore, even if in this paper we are mainly interested in superconformal theories, our results will also apply to these more general setups. 
Let us see the necessity of the CKS equation, working in Lorentzian signature for simplicity (the Euclidean case being analogous, with a doubling of the equations).
When the field theory is coupled to background new minimal supergravity, the supersymmetry condition is 
\be\label{NewMinimalEq}
\Big(\nabla_m + \frac \ii 2 v^n \gamma_{nm} + \ii\,v_m -\ii\,a_m \Big) \zeta_+ \ = \ 0\,,
\ee
where the supergravity auxiliary fields $a$ and $v$ are a gauge connection and a well-defined one-form, respectively, with the latter satisfying $\diff\!*v =0\,$. Redefining $a = A + \frac 32 v$, and contracting the equation with $\gamma^m$, one obtains $ v\cdot \zeta_+\ = \ -\frac{\ii}{2}\nabla^A\cdot \zeta_+$.
Plugging this back into \eqref{NewMinimalEq}, one arrives at the CKS equation~\eqref{ChargedCKS}. In fact, in a neighborhood where the spinor is non-vanishing, one can see that the new minimal equation~\eqref{NewMinimalEq} and the CKS equation are equivalent in Euclidean signature~\cite{KTZ,Dumitrescu:2012ha}, while they are equivalent modulo a conformal transformation in Lorentzian signature~\cite{Cassani:2012ri}.  
The condition stemming from coupling the field theory to old minimal supergravity reads
\be\label{OldMinimalEq}
\Big(\nabla_m  + \frac{\ii}{3} b_m +\frac{\ii}{6} b^n \gamma_{nm}\Big) \zeta_+ + \frac 16 M \gamma_m \widetilde \zeta_- \ = \ 0\,,
\ee
where the auxiliary fields $b$ and $M$ are a globally well-defined one-form and a complex scalar, respectively. It follows that $M\widetilde\zeta_- = - \frac 32 (\nabla - \frac{\ii}{6}b)\cdot\zeta_+$. Plugging this back into the equation and identifying $b=-2A$, again reproduces the CKS equation. However, \eqref{OldMinimalEq} is a more restrictive condition, in part because one is not allowed to gauge-transform $b$ \cite{Dumitrescu:2012at}.


Having highlighted the universality of the CKS equation as a necessary requirement for supersymmetric backgrounds, we now pass to study its integrability condition in detail. 
This is derived in appendix~\ref{DerivationIntegrability}, and reads
\be\label{IntegrabilityChargedCKS}
\Big(\frac{1}{4}C_{mnpq} -\frac \ii 3 g_{p[m} F_{n]q} \Big)\gamma^{pq} \zeta_+ - \frac \ii 3 \Big( F_{mn} - \frac 12 \gamma_{mnpq} F^{pq} \Big)\zeta_+\ = \ 0\,,
\ee
where
\be\label{defWeyl}
C_{mnpq} \ = \ R_{mnpq} -  g_{m[p} R_{q]n} + g_{n[p} R_{q]m}  + \frac 13 R\, g_{m[p} g_{q]n}\,
\ee
is the Weyl tensor of the background metric, with $R_{mnpq}$, $R_{mn}$ and $R$ denoting the Riemann tensor, Ricci tensor and Ricci scalar, respectively. 
The geometric information contained in this condition can be extracted by contracting it with a basis of chiral spinors. In the following, we study the Lorentzian and Euclidean cases in turn.
In Euclidean signature, we will also need to consider the integrability condition of the CKS equation~\eqref{ChargedCKSminus} for $\widetilde \zeta_-$. This takes the same form as~\eqref{IntegrabilityChargedCKS}, with the substitutions $\zeta_+ \to \widetilde \zeta_-$ and $F \to - F$.

\subsection{Lorentzian signature} 

In Lorentzian signature, the chiral spinor $\zeta_+$ defines an $\mathbb R^2$ structure. This is characterized by the bilinears~\cite{Cassani:2012ri}
\be \label{eq:z}
 z_m \,=\, \frac{1}{4}\, \overline{\zeta_+} \gamma_m\, \zeta_+\, , \qquad 
\omega_{mn}\, =\, -\frac{1}{4}\,  \overline{(\zeta_+)^c}\,  \gamma_{mn} \, \zeta_+ \,,
\ee
see appendix~\ref{ConventionsApp} for our spinor conventions.
By construction, the one-form $z$ is real and null, $z_m z^m = 0$, while $\omega$ is a complex two-form satisfying $z \wedge \omega =0$, which implies $\omega= z \wedge w$ for some complex one-form $w$. The one-forms $z$ and $w$ can be seen as elements of a frame, which may be completed by 
introducing an additional real one-form $e^-$ (that however does not follow from the spinor $\zeta_+$). Hence we can write $z = e^0 + e^1$, $e^- = -e^0 + e^1$, $w = e^2 - \ii e^3$. These satisfy the null tetrad relations
\be
z_m z^m = e^-_m e^{-\,m} = w_m w^m = z_m w^m = e^-_m w^m =  0\,, \qquad e^-_m z^m = w_m \bar w^m = 2\,,
\ee
and the metric reads
\be\label{eq:metric}
	\diff s^2 \,=\,  z \,  e^- + w \,  \bar w\ .
\ee
Note that we are assuming that $\zeta_+$ is nowhere vanishing, otherwise our analysis will only be valid outside its zero locus.
The two-form $\omega$ is anti-self-dual, $\ii *\omega = -\omega$.\footnote{In~\cite{Cassani:2012ri}, a form satisfying the very same property was called imaginary {\it self}-dual. Here we changed nomenclature for consistency with the Euclidean case, where the bilinears of a positive-chirality spinor turn out to be anti-self-dual (see below).}
A convenient way to parameterize the two-forms is to introduce the following anti-self-dual basis (generating the (0,1) representation of $SO(1,3)$):
\bea
\omega &=& z\wedge w\,,\nn \\[1mm]
\theta &=& e^- \wedge \bar w\,,\nn\\ [1mm]
j &=& \frac \ii 2 (z \wedge e^- + w \wedge \bar w)\,.\label{ASDbasisLorentz}
\eea
A basis of self-dual two-forms is obtained by simply complex-conjugating the anti-self-dual basis. 
 The forms~\eqref{ASDbasisLorentz} satisfy the relations
\bea\label{relationsBasisForms}
&&j_m{}^p \,j_p{}^n = -\delta_m{}^n \,,\qquad  \omega_m{}^p \omega_{pn} = \theta_m{}^p \theta_{pn} = 0  \,,\nn \\ [1mm]
&& j_m{}^p \omega_{pn}  = \,\ii\,\omega_{mn} \,,\qquad j_m{}^p \,\theta_{pn} = - \ii\,\theta_{mn} \,,\qquad \theta_m{}^p \omega_p{}^n = -2(\delta_m{}^n +\ii\, j_m{}^n)\,,\quad
\eea
which imply
\be
j_{mn} j^{mn} = 4 \,,\qquad \omega_{mn} \theta^{mn} =8  \,,\qquad   \omega_{mn} \omega^{mn} = \theta_{mn} \theta^{mn} = \omega_{mn} j^{mn} = \theta_{mn} j^{mn} = 0\,.
\ee
As for the products of a self-dual with an anti-self-dual form, for the purpose of our computation it will be sufficient to note that their anti-symmetric parts all vanish ($j_{[m}^{\;p} \,\overline j_{n]p} = 0$, etc.) and that the same is true for the contraction of both indices ($j_{mn} \,\overline j{}^{\,mn} = 0$, etc.).

In Lorentzian signature, a basis of positive-chirality spinors is given by $\zeta_+$ and $e^-_m \gamma^m (\zeta_+)^c$. Hence a basis of positive-chirality Dirac conjugates is given by $\overline{(\zeta_+)^c}$ and $\overline{\zeta_+}\,e^-_m \gamma^m$. Contracting the integrability condition~\eqref{IntegrabilityChargedCKS} with either one of these barred spinors, we obtain
\bea\label{ConsequenceIntegrab}
\Big(\frac{1}{4}C_{mnpq} -\frac \ii 3 g_{p[m} F_{n]q} \Big)\omega^{pq} & = & 0\,,\nn \\ [2mm]
\Big(\frac{1}{4}C_{mnpq} -\frac \ii 3 g_{p[m} F_{n]q} \Big)j^{pq} &=& \frac 23 F^-_{mn} \,,
\eea
where we introduced the anti-self-dual part $F^-_{mn} = \frac 12(F_{mn} - \ii \,\widetilde F_{mn})$ of the gauge field-strength, with $\widetilde F_{mn} = \frac 12 \epsilon_{mn}{}^{pq}F_{pq}\,$.

In order to efficiently work out the contractions in~\eqref{ConsequenceIntegrab}, we exploit the well-known fact that the Weyl tensor can be seen as a linear operator acting on the space of two-forms. The analysis of the algebraic properties of this operator leads to the Petrov classification, see e.g.\ Chapters 3, 4 of~\cite{ExactSolsEinstein} for an extensive discussion.
Defining the dual of the Weyl tensor as
\be
\widetilde C_{mnpq} \ =\ \frac 12 \epsilon_{mn}{}^{rs}C_{rspq}\ =\  \frac 12 \epsilon_{pq}{}^{rs}C_{mnrs}\,,
\ee
we can introduce its self-dual and anti-self-dual parts by 
\be
C^{\pm}_{mnpq}\ =\ \frac 12 (C_{mnpq} \pm \ii \,\widetilde C_{mnpq} )\,.
\ee
Note that both $C^-$ and $C^+ = \overline{C^-}$ are complex and contain the same degrees of freedom as $C$, i.e.\ ten real components. By contraction of the last pair of indices, $C^+$ acts on the self-dual two-forms, while $C^-$ acts on the anti-self-dual ones. Adopting the Newman--Penrose tetrad formalism~\cite{NewmanPenrose}, 
 we can use the basis $\omega$, $\theta$, $j$ defined above to write the anti-self-dual part of the Weyl tensor as
\bea\label{paramWeyl}
C^-_{mnpq} &=& \frac 12\Psi_0\, \theta_{mn}\theta_{pq} + \Psi_1 \left(\theta_{mn}j_{pq} + j_{mn} \theta_{pq}\right) + \frac 12\Psi_2 \left(\omega_{mn}\theta_{pq} + \theta_{mn}\omega_{pq} - 4j_{mn}j_{pq}\right)\nn \\ 
&& +\, \Psi_3 \left(\omega_{mn}j_{pq} + j_{mn}\omega_{pq}\right) + \frac 12\Psi_4\, \omega_{mn}\omega_{pq}\,,
\eea
where the Weyl scalars parameterizing the tensor components are
\bea
\Psi_0 &=& \frac{1}{32} C_{mnpq}\omega^{mn}\omega^{pq}\,,  \qquad\quad \Psi_3 \ =\ \frac{1}{32} C_{mnpq}\theta^{mn}j^{pq} \,,  \nn \\ [1mm]
\Psi_1 &=&  \frac{1}{32} C_{mnpq}\omega^{mn}j^{pq} \,,\qquad\quad\, \Psi_4 \ =\ \frac{1}{32} C_{mnpq}\theta^{mn}\theta^{pq} \,,\nn \\ [1mm]
\Psi_2 &=& \frac{1}{32} C_{mnpq}\theta^{mn}\omega^{pq} \;=\; -\frac{1}{32}C_{mnpq}j^{mn}j^{pq} \,.\label{WeylScalars}
\eea
This parameterization has the appealing feature of automatically incorporating all the algebraic symmetries of the Weyl tensor: $C_{mnpq}= C_{[mn]pq} = C_{mn[pq]} = C_{pqmn}$ and, slightly less obviously, $C^m{}_{nmp} = 0$ and $C_{m[npq]}=0\,$. 
The five Weyl scalars are complex and independent, and exactly parameterize the ten real components of the Weyl tensor.
We also record that
\be\label{C+squaredGeneralLor}
C^-_{mnpq}C^{-\,mnpq} \, \equiv \,\frac 12\big( C_{mnpq}C^{mnpq} - \ii\, C_{mnpq}\widetilde C^{mnpq}\big) \;= \; 32\left(\Psi_0\Psi_4 + 4 \Psi_1\Psi_3 + 3\Psi_2^{\,2}\right).
\ee

The forms $j$, $\omega$, $\theta$ can also be used to parameterize the gauge field $F = F^+ + F^-$:
\be
F^- \ = \ \Phi_j \,j + \Phi_\omega\, \omega + \Phi_\theta\, \theta\,,
\ee
with the $\Phi$'s being arbitrary complex functions, and $F^+$ being given by $F^+ = \overline{F^-}$.
Note that
\be\label{F+squaredGeneralLor}
F^-_{mn}F^{-mn} \ \equiv \ \frac 12 \big(F_{mn}F^{mn} - \ii\,F_{mn}\widetilde F^{mn} \big)\ = \  16 \,\Phi_\omega \Phi_\theta + 4\,\Phi_j^2\,.
\ee

Using this parameterization of the Weyl tensor and the gauge field, and recalling the relations~\eqref{relationsBasisForms}, it is straightforward to show that equations~\eqref{ConsequenceIntegrab} arising from the integrability condition of the 
CKS equation are equivalent to 
\be\label{LorentzianResult}
\Psi_0 = 0\,, \qquad \Psi_1 \,=\, \frac 13 \Phi_\theta \,, \qquad \Psi_2 \,=\, -\frac 13 \Phi_j\,,\qquad \Psi_3 \,=\,\Phi_\omega\,,
\ee
with no constraints being imposed on $\Psi_4$. Thus, the field-strength $F_{mn}$ has been determined completely in terms of the Weyl
tensor.\footnote{We believe that~\eqref{LorentzianResult}, supplemented with the existence of a null conformal Killing vector, should be equivalent to a solution of the CKS equation. However, we have not verified this.}
An immediate consequence of these relations, which is seen recalling~\eqref{C+squaredGeneralLor} and \eqref{F+squaredGeneralLor}, is
\be
C^-_{mnpq}C^{-\,mnpq} \ = \ \frac 83\, F^-_{mn}F^{-\,mn}\,.
\ee
Separating the real and imaginary parts, we obtain
\be\label{Csq=FsqLorentzian}
C_{mnpq}C^{mnpq} \ =\ \frac 83\, F_{mn}F^{mn}\,,\qquad C_{mnpq}\widetilde C^{mnpq} \ =\ \frac 83\, F_{mn}\widetilde F^{mn}\,.
\ee
It is interesting to remark that the charged conformal Killing spinor equation  (\ref{ChargedCKS}) coincides with the unbroken supersymmetry condition for bosonic backgrounds of minimal conformal supergravity, see e.g.~\cite[eq.\:(2.37)]{FradkinTseytlin}. Therefore, the first  equation in (\ref{Csq=FsqLorentzian}) shows  that on such supersymmetric backgrounds the conformal supergravity action, whose bosonic part in our conventions is given by $S \sim \int \diff^4x \sqrt{g} \left( C_{mnpq}C^{mnpq} - \frac{8}{3}F_{mn} F^{mn} \right)$, vanishes. However, note that, differently from ordinary supergravity, this supersymmetry condition, supplemented with the Maxwell equation, does not imply the equation of motion of conformal supergravity.

Vanishing of some of the Weyl scalars corresponds to different types in the Petrov algebraic classification of the Weyl tensor (see e.g.\ \cite{ExactSolsEinstein}). The generic supersymmetric background with just $\Psi_0 = 0$ is algebraically general (Petrov type~I). The property $\Psi_0 = 0$ means that our vector $z$ built from $\zeta_+$ is a principal null direction of the Weyl tensor, i.e.\ $z^n z_{[r} C_{m]np[q}z_{s]} z^p = 0$. If any Weyl scalar beyond $\Psi_0$ vanishes, then the Weyl tensor is said algebraically special; depending on which Weyl scalars are zero, it and can be of Petrov type II, III, D, N or O. In particular, when the gauge field vanishes we have $\Psi_0 = \Psi_1 = \Psi_2 = \Psi_3 = 0$ with $\Psi_4$ generically non-vanishing, which corresponds to a Petrov~N spacetime.\footnote{The uncharged  ($A=0$) CKS equation had been studied prior to~\cite{Cassani:2012ri}, in~\cite{lewandowski}. It was found that it admits solutions on Fefferman and on pp-wave-type  spacetimes (depending 
on 
whether $z$ is twisting or non-twisting, respectively). That the uncharged CKS equation implies Petrov~N can be found e.g.\ in~\cite{PenroseRindlerVol2}, around eqs.~(6.1.6) and~(8.1.4).} If in addition $\Psi_4=0$, the full Weyl tensor vanishes and the spacetime is conformally flat (Petrov type~O). Other algebraically 
special cases are also possible, for instance if $F^-$ is proportional to $\omega$ then we have $\Psi_0 = \Psi_1 = \Psi_2 = 0$, corresponding to a Petrov~III spacetime. This implies that the two sides of both equations~\eqref{Csq=FsqLorentzian} vanish separately, that is $C^2 = F^2 = C\widetilde C=F\widetilde F = 0$. 

Note that our analysis is purely local, and the Petrov type can change at different spacetime points.
We will comment more on the implications of these results in section~\ref{mainsection}.

\subsection{Euclidean signature}\label{integrabilityEuclidean}

A completely parallel analysis can be done in Euclidean signature, albeit for the self-dual and anti-self-dual parts of the Weyl and Maxwell tensors separately. This is due to the fact that contrarily to their Lorentzian analogues, Euclidean forms of opposite chiralities, defining the (1,0) and (0,1) representations of $SO(4)$, are not related by complex conjugation.

We will assume that the chiral spinor $\zeta_+$ does not vanish anywhere, being understood that if it does our local analysis will still be valid outside its zero locus. In Euclidean signature, $\zeta_+$ defines a $U(2)$ structure, and we can construct the bilinears (see e.g.~\cite{KTZ}):
\be
\label{bilibili}
\zeta_+^\dagger \zeta_+ \,=\, ||\zeta_+||^2\,,\qquad j^-_{mn} \,=\, \frac{-\ii}{||\zeta_+||^2}\,\zeta_+^\dagger\gamma_{mn}\zeta_+\,,\qquad \omega^-_{mn} \,=\,  \frac{-1}{||\zeta_+||^2}\, (\zeta_+^c)^\dagger \gamma_{mn}\zeta_+\,.
\ee
The two-form $j^-$ is real while $\omega^-$ is complex and decomposable (namely, it can be written as the wedge of two complex one-forms).
In our conventions they turn out to be anti-self-dual,
\be
*j^- \,=\, -j^-\,,\qquad *\omega^- \,=\, -\omega^-\,.
\ee
Starting from $\widetilde \zeta_-$, we can construct similar bilinears $j^+$, $\omega^+$, that are instead self-dual. The forms are related to the metric via
$g_{mn}\,=\, j^\pm_{mp} I^{\pm\,p}{}_n$, where $I^\pm$ is the almost complex structure defined by $\omega^\pm$.
Moreover, they are non-degenerate and related to the volume form as
\be
\tfrac 12 j^\pm\wedge j^\pm \,=\, \tfrac 14 \omega^\pm \wedge \overline {\omega^\pm} \,=\, \pm\, {\rm vol}_4 \,,
\ee 
with all other wedgings between them vanishing.
Calling $\theta^\pm = \overline{\omega^\pm}$, one can show that for each chirality exactly the same properties~\eqref{relationsBasisForms} found in Lorentzian signature are satisfied (this explains why we are giving the same names to the Lorentzian and Euclidean two-forms). We also note that both the commutator and the contraction of a self-dual with an anti-self-dual form vanish ($j_{[m}^{-\;p} \,j^+_{n]p} = 0$, $\;j^{-}_{mn} \,j^{+\,mn} = 0$, etc.).

A basis of positive chirality spinors is given by $\zeta_+$ together with its charge conjugate $\zeta_+^c$. Contracting the integrability condition~\eqref{IntegrabilityChargedCKS} with the hermitian conjugate of either one of these spinors, we find formally the same equations~\eqref{ConsequenceIntegrab} obtained in Lorentzian signature. If on the other hand we start from the negative chirality spinor $\widetilde \zeta_-$ and its charge conjugate, we obtain similar equations, with a few sign changes. In detail, we have
\bea\label{ConsequenceIntegrabNegChir}
\Big(\frac{1}{4}C_{mnpq} \pm\frac \ii 3 g_{p[m} F_{n]q} \Big)\omega_\pm^{pq} & = & 0\,,\nn \\ [2mm]
\Big(\frac{1}{4}C_{mnpq} \pm\frac \ii 3 g_{p[m} F_{n]q} \Big)j_\pm^{pq} &=& \mp \, \frac 23 F^\pm_{mn} \,,
\eea
where the lower sign is associated with the $\zeta_+$ equation~\eqref{ChargedCKS}, while the upper sign is associated with the $\widetilde \zeta_-$ equation~\eqref{ChargedCKSminus}. $F^\pm$ are the (anti)-self-dual parts of $F$: $F^{\pm}_{mn} = \frac 12(F_{mn}\pm \widetilde F_{mn})$.

These equations can be analysed with the same approach used in Lorentzian signature. Namely, we can introduce the Euclidean self-dual and anti-self-dual parts of the Weyl tensor
\be
C^{\pm}_{mnpq} = \frac 12 (C_{mnpq} \pm \widetilde C_{mnpq} )\,,
\ee
Note that, in contrast to the Lorentzian case, where they are conjugate to each other, in Euclidean signature $C^\pm$ are real, a priori independent tensors.
Using the anti-self-dual basis $j^-$, $\omega^-$, $\theta^-$, the anti-self-dual Weyl tensor $C^-$ can be parameterized as in~\eqref{paramWeyl}. The same can be done for the self-dual tensor $C^+$ by using the basis $j^+$, $\omega^+$, $\theta^+$. 
We thus have two sets of Weyl scalars defined as in~\eqref{WeylScalars}: $\Psi^-_0,\ldots, \Psi^-_4$ defined by means of the anti-self-dual basis, and $\Psi^+_0,\ldots, \Psi^+_4$ defined in terms of the self-dual basis.
The relations $\theta^\pm = \overline{\omega^\pm}$ and $\overline{j^\pm} = j^\pm$, together with the reality of the Weyl tensor, imply the following constraints between Weyl scalars of same chirality:
\be\label{WeylScalarsEuclidean}
\overline{\Psi_4^{\,\pm}} = \Psi_0^\pm \,,\qquad \overline{\Psi_3^{\,\pm}} = \Psi_1^\pm\,,\qquad \overline{\Psi_2^{\,\pm}} = \Psi_2^\pm\,.
\ee
We see that the $\Psi^+$ (respectively, $\Psi^-$) Weyl scalars parameterize the five degrees of freedom in the anti-self-dual (respectively, self-dual) parts of the Weyl tensor. Altogether, they parameterize the ten independent real components of the Weyl tensor.\footnote{This parameterization of the Weyl tensor in Euclidean signature was given in~\cite{Batista}, where a complex frame was used rather than the $U(2)$ structure two-forms. Since our form $\omega$ is decomposable, 
the two descriptions are locally equivalent.
}
The square of the chiral parts of the Weyl tensor read
\be\label{C+squaredGeneral}
C^\pm_{mnpq}C^\pm{}^{mnpq} \, \equiv \,\frac 12 \big( C_{mnpq}C^{mnpq} \pm C_{mnpq}\widetilde C^{mnpq}\big) \, = \, 32\left[\Psi_0^\pm\Psi_4^\pm + 4 \Psi_1^\pm\Psi_3^\pm + 3(\Psi_2^\pm)^2\right].
\ee

We can use $j^\pm$, $\omega^\pm$, $\theta^\pm$ to also parameterize the gauge field $F = F^+ + F^-$:
\be
F^\pm   \ = \ \Phi_j^\pm j^\pm + \Phi_\omega^\pm \omega^\pm + \Phi_\theta^\pm \theta^\pm\,.
\ee
Since in Euclidean signature we are allowing for a complex gauge field, the components $\Phi_j^\pm,\,\Phi_\omega^\pm,\,\Phi_\theta^\pm$ are a priori arbitrary complex functions.
Note that
\be\label{F+squaredGeneral}
F^\pm_{mn}F^{\pm\,mn} \ \equiv \ \frac 12\big( F_{mn}F^{mn} \pm F_{mn}\widetilde F^{mn}\big) \ = \ 16 \,\Phi_\omega^\pm \Phi_\theta^\pm + 4(\Phi_j^\pm)^2\,.
\ee

Applying this formalism to eq.~\eqref{ConsequenceIntegrabNegChir}, it is easy to see that the integrability condition of the CKS equation for $\zeta_+$ leads to
\be\label{ResultZeta+}
\Psi_0^- = 0\,, \qquad \Psi_1^- \,=\, \frac 13 \Phi_\theta^- \,, \qquad \Psi_2^- \,=\, -\frac 13 \Phi^-_j\,,\qquad \Psi_3^- \,=\,\Phi_\omega^-\,, 
\ee
with no constraints on the $\Psi^+$ Weyl scalars nor on the $\Phi^+$ components of $F$. Taking into account the reality conditions~\eqref{WeylScalarsEuclidean}, we arrive at
\bea\label{WeylFromIntegrabilityEuclidean}
&&\Psi_0^- = \Psi_4^- = 0\,, \qquad \overline{\Psi_1^-} \,=\, \Psi_3^-\,=\, \Phi_\omega^- \,, \qquad \Psi_2^- \,=\, -\frac 13\, {\rm Re}\,\Phi_j^-\,,\quad\nn\\ [2mm]
&&{\rm Im}\, \Phi^-_j = 0\,, \qquad  \Phi^-_\theta \,=\,3\, \overline{\Phi^-_\omega}\,.
\eea
Note that the second line imposes some ``reality'' constraints on the anti-self-dual part of the gauge field. In particular, recalling~\eqref{F+squaredGeneral} we have that although $F$ is complex, $(F^-)^2$ is real. Also recalling~\eqref{C+squaredGeneral}, we find that
\be
C^-_{mnpq}C^{-\,mnpq} \ = \ \frac 83\, F^-_{mn}F^{-\,mn}\,,
\ee
which is the same as
\be
\label{euclideanano}
C_{mnpq}C^{mnpq} - \frac{8}{3}\, F_{mn}F^{mn} \ = \ C_{mnpq}\widetilde C^{mnpq} - \frac{8}{3}\, F_{mn}\widetilde F^{mn}\,.
\ee
This is a weaker condition than the one found in Lorentzian signature, where the two sides of the equation were vanishing separately.

Similarly, integrability of the CKS equation for $\widetilde \zeta_-$, together with~\eqref{WeylScalarsEuclidean}, yields a set of conditions like~\eqref{WeylFromIntegrabilityEuclidean}, with the replacements $\Psi^- \to \Psi^+$, $\Phi^-\to - \Phi^+$, and no constraints on the $\Psi^-$ and $\Phi^-$ components. This implies
\be\label{C+Sq=F+Sq}
C^+_{mnpq}C^{+\,mnpq} \ = \ \frac 83 \,F^+_{mn}F^{+\,mn}\,,
\ee
which is the same as
\be
C_{mnpq}C^{mnpq} - \frac{8}{3}\,F_{mn}F^{mn} \ = \ -C_{mnpq}\widetilde C^{mnpq} + \frac{8}{3}\,F_{mn}\widetilde F^{mn}\,.
\ee

A classification similar to the one of Petrov exists in Euclidean signature for the self-dual and anti-self-dual parts of the Weyl tensor independently, with the only allowed types being I, D and O for each chirality, see e.g.~\cite{Batista}.
Here, referring for definiteness to a solution to the CKS equation for $\zeta_+$, we observe that the conditions~\eqref{WeylFromIntegrabilityEuclidean} imply that if $F^-$ is real (i.e.\ ${\rm Im}\,\Phi^-_j=0$, $\Phi_\theta = \overline{\Phi^-_\omega}\,$), then it can only be proportional to $j^-$, and enforces $\Psi_0^-=\Psi_1^-=\Psi_3^-=\Psi_4^- =0$, meaning that the anti-self-dual part of the Weyl tensor is of type D.
On the other hand, when $F^-$ is purely imaginary (i.e.\ ${\rm Re}\,\Phi_j=0$, $\Phi_\theta = - \overline{\Phi_\omega}\,$), all the Weyl scalars $\Psi^-$ and the gauge field components $\Phi^-$ have to vanish, hence both the Weyl and the Maxwell tensors are self-dual. Note that for a purely imaginary gauge field, if $\zeta_+$ is a solution to the CKS equation, its charge conjugate $\zeta_+^c$ is an independent solution to the same equation; hence we obtain two supercharges with the same R-charge. These observations about a purely imaginary gauge field are consistent with the results of~\cite[sect.$\:$5]{Dumitrescu:2012ha}.


\section{Taming the superconformal anomalies}

\label{mainsection}

Below we will discuss the implications of the results in the previous section on the superconformal anomaly.
For the sake of clarity we will keep the discussion of the Lorentzian and Euclidean signatures separated.

\subsection{Lorentzian signature}

Our results  establish the absence of $c$-terms in the trace and R-symmetry anomalies for $\mathcal N=1$ superconformal field theories coupled to a curved supersymmetric background.
Let us write the anomaly formulae  (\ref{correcttrace}), (\ref{correctchiral}) by separating the terms multiplied by the central charges $a$ and $c\,$:
\bea
 T_m^m   & = & \frac{c}{16\pi^2}  \Big(\weylsq  -      \frac{8}{3}   F^2 \Big) - \frac{a}{16\pi^2}  \euler     ~, \nn\\ [2mm]
\nabla_m J^m & = & \frac{c}{24\pi^2} \Big( \pontry- \frac{8}{3} F  \widetilde F \Big) + 
\frac{a}{24\pi^2} \Big( -\pontry+ \frac{40}{9} F \widetilde F \Big) \,.
 \eea 
The quadratic Weyl, Euler and Pontryagin\footnote{The density $\pontry$ is related to the first Pontryagin class of the manifold $M$ 
as $p_1(M)=  \frac{\pontry}{16\pi^2}\, \mathrm{vol} (M)$, where $\mathrm{vol}(M)$ is the volume form of $M$.} 
densities are defined as 
\bea
 \weylsq  & \equiv  &   C_{mnpq}C^{mnpq} \ = \ R_{mnpq}R^{mnpq} - 2 R_{mn}R^{mn} + \frac 13 R^2 \, , \nn\\[2mm]
  \euler  & \equiv  & \frac{1}{4}\epsilon^{mnpq}\epsilon^{rsuv}R_{mnrs}R_{pquv} \  = \ R_{mnpq}R^{mnpq} - 4 R_{mn}R^{mn} + R^2\, , \nn \\[2mm]
  \pontry & \equiv & \frac{1}{2}\epsilon^{mnpq} R_{mnrs}R_{pq}{}^{rs} \ = \ \frac{1}{2}\epsilon^{mnpq} C_{mnrs}C_{pq}{}^{rs}\label{definvariants}
  \, ,
\eea
where the last equalities in the first and in the third lines follow straightforwardly from the definition~\eqref{defWeyl} of the Weyl tensor.
We have also introduced the short-hand notation $F^2\equiv F_{mn}F^{mn}$ and $F\widetilde F \equiv\frac{1}{2}\epsilon^{mnpq} F_{mn} F_{pq}$.
The conditions  (\ref{Csq=FsqLorentzian}) show that for supersymmetric field theories in supersymmetric backgrounds 
\emph{the terms multiplied by $c$ vanish in the trace and R-current anomalies}, hence both anomalies are purely topological invariants.\footnote{In Lorentzian signature, the manifolds of physical relevance 
are non-compact, hence defining topological invariants is more subtle than on compact spaces. Here we will refrain from integrating the anomalies.}
Specifically, the anomaly formulae reduce to  
\bea
 T_m^m  & = &  - \frac{a}{16\pi^2}  \euler     ~, \nn\\ [2mm]
\nabla_m J^m & = &  \frac{a}{36\pi^2} \pontry  \ = \  \frac{2\, a }{27\pi^2}F \widetilde F  ~.
\label{lorentzanomalies}
 \eea
 
In~\cite{Cassani:2012ri} it was shown that a necessary and sufficient condition for the existence of a solution to the CKS 
equation is that there exists a null conformal Killing vector, with a natural coordinate system adapted to this. 
It is straightforward to compute the Weyl scalars $\Psi_0, \dots, \Psi_4$ in the explicit coordinates given in \cite{Cassani:2012ri}, and obtain $F$ using the relations (\ref{LorentzianResult}). 
In particular,  computing the  Euler and Pontryagin densities explicitly, we have checked that generically they do not vanish. 
However, in certain classes of geometries the Pontryagin density does vanish, implying that 
the $U(1)_R$ current is not anomalous on these backgrounds. The following are sufficient criteria for the vanishing of $\pontry$: 
\begin{itemize}
 \item Using (\ref{C+squaredGeneralLor}) we have
 \bea
 \pontry & = & - 64\, \mathrm{Im} \left( 4 \Psi_1 \Psi_3 + 3 \Psi_2^2 \right)\,,
 \eea
which  vanishes for space-times of Petrov type O (where $C_{mnpq}=0$), N (where $F_{mn}=0$), and III (where $\weylsq =0 =\pontry$).  An example of Petrov type III is the 
$SU(2)\times U(1)$-invariant background on $\mathbb{R} \times S^3$, arising as the boundary of a deformation of AdS$_5$, discussed in \cite{GGS}. Due to the split topology of the space and the homogeneity of the metric, in this specific example one also has $\euler =0$, so the full superconformal anomaly vanishes.

 \item If there exists a (conformal) Killing vector $k$ that is also hypersurface-orthogonal, namely the dual one-form  satisfies $k \wedge \diff k=0$, then $\pontry=0$. 
 This may be verified noting that since $\pontry$ is a conformal invariant density,
 in the conformal class of metrics we can consider one where $k$ is a Killing vector. 
 If the Killing vector  $k=\de_t$ is time-like or space-like, then the metric is a Riemannian product 
 $\diff s^2 =  \pm \diff t^2 + \diff s^2 (M_3)$,  where $M_3$ is a 
 three-dimensional manifold.\footnote{Note that the existence of a time-like hypersurface-orthogonal Killing vector is also a sufficient condition for Wick-rotating the background to Euclidean signature.}
 By direct computation of the curvature one can verify that $\pontry =0$. 
 If $k$ is null, then this corresponds to the case of non-twisting geometries discussed in \cite{Cassani:2012ri}, and again a direct computation shows that $\pontry =0 $. 
An example possessing such Killing vectors is given by the boundary of the five-dimensional magnetic string solutions of \cite{ChamseddineSabra,KlemmSabra}. These comprise a metric on $\mathbb{R}^{1,1}\times S^2$ or $\mathbb{R}^{1,1}\times \mathbb{H}^2$ of Petrov type~D (which has all Weyl scalars but $\Psi_2$ equal zero), and a non-trivial gauge field. Moreover, on this background $\euler =0$, so the superconformal anomaly fully vanishes. Another Petrov D example with similar properties will be discussed later in section~\ref{theexample}.

\end{itemize}

\subsection{Euclidean signature}\label{EuclideanComments}

In Euclidean signature, the integrability  equation (\ref{IntegrabilityChargedCKS}) for $\zeta_+$ does not contain informations about the 
self-dual parts of the Weyl and Maxwell tensors. 
Although these are completely determined by the CKS equation \cite{Dumitrescu:2012ha, KTZ}, 
the weaker condition (\ref{euclideanano}) is sufficient to show that the $c$-contribution to the conformal anomaly, 
on the left hand side of (\ref{euclideanano}), is a topological density, on the right hand side of (\ref{euclideanano}). 
Therefore, as in the Lorentzian case, the trace anomaly is purely topological, although generically the  $c$-contribution to the anomalies does not vanish. 
Namely, the anomalies read
\bea
\label{integerrimo}
 T_m^m   & = & \frac{c}{16\pi^2}  \Big[\pontry - \frac{8}{3} \mathrm{Re} (F \widetilde F) \Big] - 
\frac{a}{16\pi^2}  \euler \,  - \mathrm{i} \, \frac{c}{6\pi^2}  \mathrm{Im} (F \widetilde F)  \,   , \nn\\[2mm]
\nabla_m J^m & = &  \frac{c-a}{24\pi^2} \pontry + \frac{5a-3c}{27\pi^2} \mathrm{Re} (F \widetilde F)  + \mathrm{i} \, \frac{5a-3c}{27\pi^2} \mathrm{Im}  (F \widetilde F) ~.
\label{lachirale}
 \eea
Although there is no new information on the R-current  anomaly,  we have emphasized that generically both  anomalies include an \emph{imaginary} part. 

On a compact Riemannian manifold $M$ without boundary, it is interesting to consider the integrated version of eqs.~(\ref{lachirale}), and express these 
in terms of the \emph{Euler characteristic} $\chi$ and \emph{signature} $\sigma$ of $M$, given by
\bea
\mathbb{Z} \ \,\ni\,   \chi (M) & = & \frac{1}{32\pi^2}\int_M \diff^4 x \sqrt{g}\, \euler  ~,\\ [2mm]
\mathbb{Z} \ \,\ni\,  \sigma (M) & = &  \frac{1}{3} \int_M p_1(M) \  = \ \frac{1}{48\pi^2} \int_M \diff^4 x \sqrt{g} \, \pontry ~.
\eea
Introducing the real and imaginary parts of the gauge field as\footnote{The factor of $\frac{1}{2}$ is introduced for convenience, as it will become clear momentarily.}
\bea
F   &= & \frac{1}{2}({\cal F} + \mathrm{i}\,  \imf)~,
\label{looksweird}
\eea
we also define
\bea
\nu & \equiv & \frac{1}{2\pi^2}\int_M \diff^4 x \sqrt{g}\, \, F \widetilde F  \ = \nn\\ [2mm]
 & = &  \int_M \frac{{\cal F}}{2\pi}  \wedge \frac{{\cal F}}{2\pi}  -
\int_M \frac{\imf}{2\pi}  \wedge \frac{\imf}{2\pi}   +  2\mathrm{i} \int_M \frac{{\cal F}}{2\pi}  \wedge \frac{\imf}{2\pi}
~.
\label{definsta}
\eea

In order to evaluate the integrals in (\ref{definsta}), we pause to recall some of the \emph{global} properties of the backgrounds, discussed in \cite{Dumitrescu:2012ha,KTZ}. Here, it is important to assume that the spinor $\zeta_+$ is well-defined and non-vanishing everywhere on $M$.
Firstly, notice that our gauge field $A$ is related to the gauge field in \cite{Dumitrescu:2012ha} as 
$A^\mathrm{here}=A^\mathrm{DFS}- \frac{3}{2}V^{\rm DFS}$, where $A^\mathrm{DFS}$ and $V^{\rm DFS}$ are the auxiliary fields of new minimal supergravity (called $a$ and $v$ at the beginning 
of section~\ref{ckseq}). However, assuming that $V^{\rm DFS}$ is a globally defined one-form on $M$, this does not affect the 
discussion of the global properties of $A$. In particular, using eq.~(3.17) in \cite{Dumitrescu:2012ha} we have (in our notation):
\bea
A_m & = & A_m^c + \frac{1}{4}(\delta_m^n + 2\mathrm{i}\, {j^-}_m{}^n)\nabla_p\, {j^-}^p{}_n \,,
\eea
where ${j^-}_m{}^n$ is the integrable complex structure on $M$, obtained raising an index on the two-form defined in (\ref{bilibili}), and
\bea
\label{defac}
A^c & = & - \frac{\mathrm{i}}{4}(\de - \bar \de ) \log \sqrt{g} - \frac{\mathrm{i}}{2}\, \diff  \log s~.
\eea
Here $s$ is a complex function that may be defined as $\zeta_+ = \sqrt{s} \,\zeta_+^0$, where $\zeta_+^0$ is a constant chiral spinor. 
The important point to note is that $A_m-A^c_m$ is a global one-form on $M$, therefore $\imf $ is an exact two-form, and 
it drops out of (\ref{definsta}). The real part of $A_m$, however, transforms as a gauge field for 
the local $U (1)_R$ symmetry,  being a connection on the line bundle ${\cal L}= \bar{\cal K}^{-1/2}$, where $\bar{\cal K} = \Lambda^{(0,2)}(M)$ is the anti-canonical bundle of $(0,2)$-forms on $M$. 
Correspondingly, the spinor $\zeta_+$ is a section of the spin$^c$ bundle ${\cal V} = \bar{\cal K}^{-1/2}\otimes {\cal S}_+$, where 
${\cal S}_+$, is the bundle of positive-chirality spinors. 
Although  neither bundle may exist separately due to the second Stiefel--Whitney class $w_2 (M) \in H_2 (M; \mathbb{Z}_2 )$ being non-trivial, the tensor product ${\cal V}$ always exists as a genuine complex vector bundle. 

Recall that on any Hermitian manifold, the  metric $g_{mn}$ and complex structure $(j^-)_m{}^n$ induce a connection 
on the anti-canonical bundle $\bar{\cal K}$, given by 
\bea
\rho & = & -\frac{\mathrm{i}}{2} (\de - \bar \de) \log \sqrt{g}~,
\eea
whose curvature is the Ricci two-form ${\cal R}_{mn} =  \frac{1}{2} (j^-)^{pq} R_{pqmn}\,$. 
The latter defines the first Chern class  $c_1(M) = \left[ \frac{{\cal R}}{2\pi} \right] \in H^2(M;\mathbb{R})$.
Comparing this with the definition of $A^c$ in (\ref{defac}), we have the following chain of identities: 
\bea
\frac{1}{2} \left[ \frac{{\cal F}}{2\pi}\right] \ = \
\left[ \frac{\diff A^c}{2\pi}\right] \  = \ 
\frac{1}{2} \left[ \frac{\diff \rho}{2\pi} \right]  \ = \ \frac{1}{2} \left[\frac{{\cal R}}{2\pi}\right]~.
\label{chain}
\eea
This is of course consistent with the fact that $2\cdot $Re$A$ is a connection on the anti-canonical bundle, and justifies the factor of $1/2$ in the definition (\ref{looksweird}). In particular, the 
 quantization condition
\bea
\int_{\Sigma_a} \frac{{\cal F}}{2\pi} &  \in &\mathbb{Z}~,
\eea
where $\Sigma_a$ is a basis of generators of $H_2(M;\mathbb{Z})$, is automatically satisfied. 
Note that on \emph{K\"ahler} manifolds the identities in (\ref{chain}) hold more strongly as identities of two-forms, rather than in co-homology.
Using  these, we  see that  the integral defining $\nu$ is the self-intersection of the first Chern class 
\bea
\mathbb{N} \, \ni \, \nu (M) &  = & \frac{1}{(2\pi)^2}\int_M {\cal F}  \wedge {\cal F}  \ = \ \int_M c_1 (M)  \wedge c_1(M)~, 
\eea
which on a complex manifold is given by
\be\label{relnusigmachi}
\nu (M) \,=\, 3 \sigma (M) + 2 \chi (M)\,.
\ee
Therefore the integrated anomalies are determined completely by the signature and Euler characteristic of $M$
 and read 
\bea
 \int_M \diff^4 x \sqrt{g}\,  T_m^m   & = &    3 c \,\sigma (M) - 2a \, \chi (M) -  \frac{c}{3}\,  \nu(M) ~, \nn\\
 \int_M \diff^4 x \sqrt{g}\, \nabla_m J^m  & =  &   2(c-a)\, \sigma (M)   + \frac{2}{27} (5a-3c)\,  \nu(M) ~.
\eea

In general, all the topological invariants are non-zero, implying that the corresponding densities are non-trivial. 
Simple examples are provided by K\"ahler manifolds, as we will discuss momentarily. 
Sufficient criteria for the vanishing of the Pontryagin density $\pontry $
are analogous to the ones discussed in Lorentzian signature. 
Moreover, we note that if there exists a non-trivial\footnote{Not proportional to the metric.} Codazzi tensor $B_{mn}$, 
\emph{i.e.} a symmetric tensor satisfying the equation $\nabla_{[m}B_{n]p}=0$,  then the Pontryagin density vanishes by Theorem 2 of 
\cite{codazzi}.\footnote{Reference \cite{codazzi} discusses Riemannian manifolds, however this criterion may as well remain valid in Lorentzian signature.}
A necessary and sufficient condition for the signature $\sigma(M)$ to be zero is that the complex manifold $M$ is the boundary of an oriented five-dimensional manifold $M_5$: $M=\de M_5$. This follows from a theorem by Rohlin (recalling that any complex manifold is oriented). In particular, the signature vanishes if the manifold is topologically $S^1\times M_3$, even if the metric is not the direct-product metric. Indeed, this is the boundary of $D^1 \times M_3$, where $D^1$ is a disk. Note that $\sigma (M)=0$ is needed in order to construct a five-dimensional holographic dual solution on a manifold $M_5$ ``filling in'' the background $M=\de M_5$.

\subsubsection*{Manifolds admitting two CKS of opposite R-charge}

When there exist two supercharges of opposite R-charge, namely a solution $\zeta_+$ to the CKS equation~\eqref{ChargedCKS} together with a solution $\widetilde \zeta_-$ to eq.~\eqref{ChargedCKSminus}, the implications on the anomalies are stronger, and analogous to the 
Lorentzian signature case. In the following, we assume that $\zeta_+$ and $\widetilde \zeta_-$ are nowhere vanishing.
This case was analysed in \cite{Dumitrescu:2012ha,KTZ}, where it was shown that there exist two (generically) 
commuting Killing vectors, constraining the form of the metric to a $\mathbb{T}^2$ fibered over a Riemann surface $\Sigma$.  Namely, in complex coordinates adapted to the complex structure 
$j^-$, this takes the form
\bea
\diff s^2 & = & \Omega^2 (z,\bar z) \left[ (\diff w + h (z,\bar z)\diff z) (\diff\bar w + \bar h (z,\bar z)\diff \bar z)  + c^2(z,\bar z) \diff z \diff \bar z\right] ~.
\label{twochargesmetrics}
\eea
Here we note that two solutions to the CKS equations of opposite chiralities
impose the constraints we discussed above, for \emph{both} self-dual and anti-self-dual parts of the Maxwell and Weyl tensors. 
In particular, we have 
\bea
\weylsq - \frac{8}{3}\mathrm{Re}\, F^2  &  = &  0 ~, \nn\\ 
\pontry  -  \frac{8}{3}\mathrm{Re}\, F \widetilde F     & = & 0 ~, \nn\\
\mathrm{Im}\, F^2 \ = \ \mathrm{Im}\, F \widetilde F  &  = &  0 ~.
\eea
These imply that the anomalies take the same form as in the Lorentzian case, namely
(\ref{lorentzanomalies}),
 and are automatically \emph{real}. Integrating on a compact manifold  without boundary,  gives the simple relations
\bea
\int_M \diff^4 x \sqrt{g}\,  T_m^m   & = &  -  2 a \, \chi (M) ~, \nn\\ [2mm]
\int_M \diff^4 x \sqrt{g}\, \nabla_m J^m & = &  \frac{4 }{3} a \, \sigma (M) ~,
\eea
as well as  the topological relation $ \sigma (M)  =   \frac{1}{9} \nu(M)$.
By evaluating the Ricci form of the metric (\ref{twochargesmetrics}),
\bea
{\cal R} & = & \mathrm{i}\, \de \bar \de \log [ \Omega^2 (z,\bar z)c (z,\bar z ) ] ~,
\eea
it is straightforward to see that this is basic with respect to the (complex) Killing vector $\de_w$, namely  $\de_w \lrcorner {\cal R}=0$. It immediately follows that $\nu (M)$ vanishes. So, recalling~\eqref{relnusigmachi}, on backgrounds preserving two supercharges of opposite R-charges we have
\bea
\sigma (M) & = & \nu(M) \ = \  \chi (M) \ =  0 ~,
\eea
hence both the integrated anomalies vanish.

\subsubsection*{Examples}

We conclude this section discussing briefly simple examples of complex manifolds, which may be used as supersymmetric backgrounds. Note that although the manifolds below
all admit K\"ahler metrics (in some cases  there exist also Einstein metrics), one is not restricted to these. The considerations we make below are purely topological.

\begin{itemize}

\item A class of examples is provided by $M=dP_k$ the $k$th \emph{del Pezzo} surface, with $k=0,\dots,8$, obtained  blowing up  $k$  points at generic positions in
 $\mathbb{C}P^2$. In particular, $dP_0=\mathbb{C}P^2$.  For $k=1,2,3$ these are toric manifolds, and  for all $k\neq 1,2$ there exist 
K\"ahler--Einstein metrics, although for $k\neq 0$ these are not known in explicit form.  The non-zero Betti numbers are  $b_0=b_4=1$, and $b_2=k+1$, 
thus the Euler characteristic is $\chi (dP_k)=k+3$. Using 
$K^2=K \cdot K = 9 - k$, where $K$ is the anti-canonical class, we obtain $\nu (dP_k ) =  9-k$ and,  by eq.~\eqref{relnusigmachi},  $\,\sigma (dP_k) = 1-k$. Inserting these expressions in the formulae for the anomalies we see that for generic values of the central charges $a$ and $c$, and for all values of $k$, these are different from zero. Note that the only del Pezzo surface having vanishing signature, and so the only one arising as the boundary of a five-dimensional manifold, is $dP_1$. Explicit Hermitian metrics on $dP_1$ were presented in \cite{Page:1979zv} and  \cite{Martelli:2004wu}.

\item Another class of simple examples is given by the product of two Riemann surfaces $\Sigma_1\times \Sigma_2$. The signature is zero since any oriented Riemann surface bounds an oriented 3-manifold $N_3$ (called handlebody), and so 
$\Sigma_1\times \Sigma_2 = \de (N_3\times \Sigma_2)$. Hence $\sigma (\Sigma_1\times \Sigma_2 )= 0$. The Euler characteristic is given by the product of the Euler characteristics of the two surfaces, hence
 $\chi (\Sigma_1\times \Sigma_2 ) = 2 (1-g_1)\cdot 2 (1-g_2)$, where $g_1,g_2$ are their genera. By formula~\eqref{relnusigmachi}, we have $\nu(\Sigma_1\times \Sigma_2 ) = 2\chi(\Sigma_1\times \Sigma_2 )$.
 A special case is that of $\Sigma_1\times \Sigma_2 = \mathbb{T}^2  \times \Sigma_2$, which is an example
admitting two supercharges of opposite R-charges. As we discussed above, these necessarily have $\sigma (M)=\chi (M)=0$.

\end{itemize}

\section{A background with vanishing anomalies}

\label{theexample}

In section~\ref{mainsection} we mentioned a few non conformally-flat backgrounds where the full trace and R-symmetry anomalies vanish, which also have a known gravity dual~\cite{GGS,ChamseddineSabra,KlemmSabra}. In this section, we present in some detail another example of this type, whose gravity dual will be the object of a separate publication. This example has both a Lorentzian and an Euclidean avatar (as~\cite{ChamseddineSabra,KlemmSabra} but differently from~\cite{GGS}, which appears to be just Lorentzian), and will also serve as an illustration of the general results in the previous sections. We focus on its Euclidean version, commenting on the Lorentzian case at the end.

We consider an $S^3\times S^1$ topology, and impose that $ SU(2)_{\rm left}\times U(1)_{\rm right}\times  U(1)_t$ symmetry is preserved. Then the metric reads
\be\label{metricS3S1}
\diff s^2 \ = \ \frac{\rS^2}{4}\left(\sigma_1^2 + \sigma_2^2 + 4s^2 \sigma_3^2\right) + \left( \diff t + k\, \sigma_3 \right)^2
\ee
and the general form of the gauge potential is
\be
A \ = \ p\, \diff t  +  q\, \sigma_3\,.
\ee
Here, the $\sigma$'s are $SU(2)$ left-invariant one-forms on $S^3\,$ parameterized by Euler angles $\vartheta, \phi,\psi$:
\bea
\sigma_1 &=& -\sin\psi \,\diff \vartheta + \cos\psi \sin\vartheta \,\diff \phi\,,\nn\\ [1mm]
\sigma_2 &=& \cos\psi\, \diff \vartheta + \sin\psi \sin\vartheta \,\diff \phi\,,\nn\\ [1mm]
\sigma_3 &=& \diff \psi + \cos\vartheta \,\diff  \phi\,,
\eea
$\rS$ is the radius of $S^3$, $s$ is a squashing parameter, and we are fixing the modulus of the $g_{tt}$ metric component to one (which can always be done by rescaling $k$ and $t$). In Euclidean signature, we allow $p$ and $q$ to be complex parameters, while the parameters in the metric are assumed real. We will work in the frame
\be\label{frameSU2U1U1}
e^1 \,=\, \frac{\rS}{2} \sigma_1\,, \quad e^2 \,=\, \frac{\rS}{2} \sigma_2 \,, \quad e^3 \,=\, \rS\,s\,\sigma_3\,,\quad e^4 \,=\,  \diff t + k\, \sigma_3\,.
\ee


We observe that the terms proportional to $k$ in the metric (\ref{metricS3S1}) can be removed by performing the change of coordinates
\be\label{changecoordspars1}
\psi' \ =\ \psi + \frac{k}{s^2\,\rS^2 + k^2}\,t\,,\qquad t' \ = \ \frac{1}{\lambda}\,t\,,
\ee
accompanied by the redefinition of the parameters
\be\label{changecoordspars2}
s'\ =\ \lambda\,s\,, \qquad p'\ = \ \lambda\,p - \frac{k}{s^2\,\rS^2 \lambda}\,q\,, 
\ee
where $\lambda = \sqrt{1+\frac{k^2}{s^2\,\rS^2}}\,$. One can check that in the primed variables the metric takes the direct product form 
\be
\diff s^2 \ = \ \frac{\rS^2}{4}\left(\sigma_1'^{\,2} + \sigma_2'^{\,2} + 4s'^{\,2} \sigma_3'^{\,2}\right) +  \diff t'^{\,2}
\ee
and the gauge field is still in the original form
\be
A \ =\ p' \diff t + q\, \sigma_3'\,.
\ee
Here, $\sigma'_1$, $\sigma'_2$, $\sigma'_3$ are $SU(2)$ left-invariant one-forms constructed using $\psi'$ at the place of $\psi$.

For any value of the radius $\rS$ and squashing $s$, the metric~\eqref{metricS3S1} allows for solutions to the CKS equation and thus yields a good supersymmetric background, provided an appropriate gauge field is chosen. Indeed, the metric describes a torus fibration over $S^2$ (with an $S^1$ in the torus being the Hopf fiber and the other being generated by $\partial_t$), hence recalling the results in~\cite{Dumitrescu:2012ha} the background allows for at least two solutions to the CKS equation with opposite R-charge.
 The gauge field can be determined using the general formulae given in~\cite{KTZ,Dumitrescu:2012ha}. Here, we apply our alternative method exploiting the integrability condition. This completely determines the field-strength $F$, and thus the potential $A$ modulo a closed one-form.

Before presenting the details, let us observe that both the Euler and Pontryagin densities $\euler$ and $\pontry$ associated with the metric~\eqref{metricS3S1} vanish. This can be checked by evaluating~\eqref{definvariants}. Another way to see it is first to observe that since~\eqref{metricS3S1} is a left-invariant metric on a homogeneous space, both $\euler$ and $\pontry$ must be constant, and then to recall that the respective integrals have to vanish, as noted in section~\ref{EuclideanComments} while discussing backgrounds with two CKS of opposite R-charge. Together with the relation between Weyl and Maxwell tensors imposed by supersymmetry, this shows that the whole Weyl and R-symmetry anomalies vanish.

\subsubsection*{Four supercharges}

Let us first review the case in which the gauge field-strength vanishes, which was already studied in~\cite{ImamuraYokoyama,Dumitrescu:2012ha}. From the analysis in section~\ref{integrabilityEuclidean}, we know that when there are two conformal Killing spinors of opposite chirality and $F=0$, then the Weyl tensor has to vanish, meaning that the space is conformally flat. For the metric~\eqref{metricS3S1}, this translates into
\be\label{ConfFlatnessCond}
k^2\ = \ \frac{\rS^2}{4} (1-4s^2) \,,
\ee
hence (choosing the positive root for $k$)
\be
ds^2 \ =\ \frac{\rS^2}{4}\left(\sigma_1^2 + \sigma_2^2 + 4s^2 \,\sigma_3^2\right) + \left(\diff t + \frac{\rS}{2}\sqrt{1-4s^2}\, \sigma_3\right)^2 .
\label{4dmetricIY}
\ee
Here, the squashing is restricted to $0<s\leq \frac 12$, namely the size of the $S^1$ Hopf fiber inside $S^3$ is not larger than the one of the base $S^2$. Having satisfied the integrability condition, one can solve the CKS equation explicitly. 
If we require the spinors not to depend on $t$, so that they are well-defined on $S^1$, then we need to introduce a purely imaginary Wilson line gauge potential
\be\label{4dgaugefieldIY}
A \ =\ \frac{\ii s}{\rS} \, \diff t  \,.
\ee
The solution for $\zeta_+$ is
\be\label{solSpinorIY}
\zeta_+ \ = \ \left(\begin{array}{cc} \cos\frac\vartheta 2 \,e^{-\frac{\ii}{2} (\psi+\phi)} & -\sin\frac\vartheta 2 \,e^{-\frac \ii 2 (\psi-\phi)} \\ [2mm]
 \mu\sin\frac\vartheta 2 \,e^{\frac \ii 2 (\psi-\phi)}  & \mu\cos\frac\vartheta 2 \,e^{\frac \ii 2 (\psi+\phi)} \end{array}\right)\zeta_+^0\,,
\ee
where $\mu \ =\  2s -\ii \sqrt{1-4s^2}$,
and $\zeta_+^0$ is any constant spinor of positive chirality. Hence we have two independent solutions to the CKS equation~\eqref{ChargedCKS}.\footnote{The new minimal equation~\eqref{NewMinimalEq} is also satisfied, with background fields
$a \ = \ v \ =\ -\frac{2\ii s}{\rS} \, \diff t\,$.} This was expected from the fact that the integrability condition has not imposed any algebraic restrictions on the spinors. On the same background, the $\widetilde\zeta_-$ equation~\eqref{ChargedCKSminus} is solved by negative-chirality spinors having the same form as~\eqref{solSpinorIY}, with $\mu$ replaced by $\bar \mu$. So any such conformally-flat background allows two independent solutions to the CKS equation for each choice of chirality, thus preserving a total of four supercharges.

When $s= \frac 12$, the metric becomes the standard, round one on the direct product $S^3 \times S^1$:\footnote{For $s= \frac 12$, there is also a solution having an opposite gauge potential, $A = -\frac{\ii}{2\rS} \diff t\,,$ and any constant $\zeta_+$ or $\widetilde \zeta_-$ being allowed; these are $SU(2)_{\rm left}$ invariant spinors on the round $S^3$, each transforming as a doublet under $SU(2)_{\rm right}$. On the other hand, spinors of the form~\eqref{solSpinorIY} are $SU(2)_{\rm right}$ invariant, and transform as a doublet under $SU(2)_{\rm left}\,$.
}
\be\label{RoundMetricS3S1}
ds^2 \ = \ \frac{\rS^2}{4}\left(\sigma_1^2 + \sigma_2^2 + \sigma_3^2\right) + \diff t^2\,.
\ee
Actually, for any allowed value of $s$ the metric \eqref{4dmetricIY} is related to~\eqref{RoundMetricS3S1} by the transformation~\eqref{changecoordspars1}, \eqref{changecoordspars2}. Indeed, specializing to $k = \frac{\rS}{2}\sqrt{1-4s^2}$, $p = \frac{\ii\,s}{\rS}$, $q=0$, the transformation reads
\be
\psi' \ = \ \psi + \frac{2}{\rS}\sqrt{1-4s^2}\:t\,,\qquad t' \ = \ 2s \,t \,,\qquad s' =\frac{1}{2}\,,\qquad p'\ = \ \frac{i}{2\rS}\,.
\ee
Since $s' = \frac{1}{2}$, we obtain the round $S^3$ metric on the direct product $S^3\times S^1$. 
This was expected from the classification of solutions to the new minimal equation given in~\cite{Dumitrescu:2012ha}, where it was found that the only compact background admitting two supercharges with same chirality and including an $S^3$ topology is (a discrete quotient of) the direct product $S^3\times S^1$, with the round metric on~$S^3$. 

Finally, we consider a reduction of this background to three dimensions. Starting from the metric~\eqref{4dmetricIY} and reducing along~$\partial_t$, we obtain the squashed $S^3$ background of~\cite{ImamuraYokoyama}. From the arguments above it is also clear that the same three-dimensional background can equally well be obtained starting from the round metric~\eqref{RoundMetricS3S1} and reducing along a $U(1)$ generated by a combination of $\partial_\psi$ and $\partial_t\,$~\cite{ImamuraYokoyama}. On the other hand, reducing~\eqref{RoundMetricS3S1} along the trivial direction $\partial_t$ leads to a round $S^3$ in three dimensions. Hence we see that different three-dimensional backgrounds can be obtained starting from the same four-dimensional configuration and reducing along different directions.

\subsubsection*{Two supercharges}

Let us now assume that $F$ is not zero. In order to evaluate the constraints \eqref{WeylFromIntegrabilityEuclidean} following from the integrability condition of the CKS equation for $\zeta_+$, and the analogous ones from the $\widetilde \zeta_-$ equation, as a first thing we construct a basis of two-forms $j^\pm,\omega^\pm,\theta^\pm$ from generic chiral spinors $\zeta_+$, $\widetilde\zeta_-$ as described in section~\ref{integrabilityEuclidean}. Then, studying the constraints involving exclusively the gauge field components $\Phi$, we find that the imaginary part of $F$ has to vanish, and also obtain some algebraic constraints on the two-forms, which can be rephrased as constraints on the spinors. Taking these into account, the only non-vanishing Weyl scalars are $\Psi_2^- = \Psi_2^+ = \frac{4k^2-\rS^2(1-4s^2)}{3\rS^4}$. The conditions relating the Weyl and Maxwell tensors then fix the real part of $F$. 
In formulae, we find that the integrability condition of the CKS equations for $\zeta_+$ and $\widetilde \zeta_-$ is solved by
\be
F = \pm\left( \frac{4s^2-1}{2} + \frac{2k^2}{\rS^2} \right)\sigma_1\wedge \sigma_2\,, \qquad \gamma^{12} \zeta_+ \,=\, \mp \ii \, \zeta_+\,,\qquad \gamma^{12} \widetilde\zeta_- \,=\, \pm \ii \, \widetilde\zeta_-\,,
\ee
where either the upper or the lower signs have to be chosen, and $k^2\ \neq \ \frac{\rS^2}{4} (1-4s^2)$. Note that the projections being imposed on the chiral spinors imply that the background admits a single solution $\zeta_+$ and a single solution $\widetilde\zeta_-$ to the respective CKS equations, meaning that exactly two supercharges are preserved.
To complete the solution, we look directly at the CKS equation, and find that any constant chiral spinor satisfying the projections above is a solution, provided one takes\footnote{In this case the new minimal equation~\eqref{NewMinimalEq} is solved by taking 
$v \ =\ \frac{2\ii s}{\rS}\diff t $ and $a= A+\frac 32v$.}
\be
A 
= \pm\left( \frac{4s^2-1}{2} + \frac{2k^2}{\rS^2} \right)\sigma_3 + \left(\pm\frac{2k}{\rS^2}  - \frac{\ii s}{\rS} \right) \diff t\,.
\label{4dgaugefieldHHL}
\ee

This supersymmetric background was already mentioned (for $k = 0$) in \cite{KTZ},
where it was identified with the lift of the three-dimensional ${SU(2)\times U(1)}$ invariant background of~\cite{HHL}. 
Of course, the change of coordinates (\ref{changecoordspars1}) and redefinition of parameters (\ref{changecoordspars2}) can be used to set $k = 0$.
However, the three-dimensional background obtained by reducing along $\partial_t$  with generic parameters $s$ and $k$ is  different from that of \cite{HHL}. 
 
\subsubsection*{Lorentzian signature}

Finally, let us briefly comment on the background in Lorentzian signature.
In this case, the topology is $S^3\times \mathbb R\,$. The Wick rotation of the metric~\eqref{metricS3S1} is performed by setting $t_{\rm E} = -\ii \,t_{\rm L}$, $k_{\rm E} = -\ii \,k_{\rm L}$. Using the results of~\cite{Cassani:2012ri}, it is immediate to see that supersymmetry is preserved because there is a globally-defined null Killing vector given by a combination of the time-like vector $\partial_t$ and the space-like vector $\partial_\psi$. Then one can follow the same steps as in the Euclidean case, implementing the analytic continuation $t_{\rm E} = -\ii \,t_{\rm L}$, $k_{\rm E} = -\ii \,k_{\rm L}$, $p_{\rm E} = \ii \,p_{\rm L}$ everywhere in the equations above. Note that in this way the gauge potential $A$ becomes real, as it has to be in Lorentzian signature.\footnote{To determine $A$ one could also use the general formulae given in~\cite{Cassani:2012ri}.
} Also, in Lorentzian signature $\widetilde \zeta_-$ is the charge conjugate of $\zeta_+$.
When $F=0$, the spacetime is conformally flat and the CKS equation has two independent solutions (four real supercharges). 
When $F\neq 0$, the only non-vanishing Weyl scalar is $\Psi_2$, hence our spacetime is of Petrov type~D and we have one solution to the CKS equation (two real supercharges). As in the Euclidean setup, the topological densities $\euler$ and $\pontry$ vanish, so there are no trace and R-symmetry anomalies.

\section{Holographic superconformal anomalies}
\label{holosection}

Our discussion so far was valid for general supersymmetric field theories,  whereas in this section we will discuss the special situation 
 when a SCFT has an asymptotically locally AdS gravity dual.\footnote{For definiteness, throughout this section we will work in Lorentzian signature.}
Although the formulae we report below are review of known results, we emphasise that the agreement of the holographic computations with the combined equations 
(\ref{correcttrace}) and (\ref{correctchiral}) was not noted before.

It is well-known that  the trace anomaly of a CFT (not necessarily supersymmetric) 
 is related to a logarithmic\footnote{In a specific coordinate system.} divergence of the on-shell bulk gravity action, and
can be obtained using the method of holographic renormalisation~\cite{Henningson:1998gx}.
On the other hand, the chiral anomaly of an axial current arises from  a Chern--Simons term~\cite{Witten:1998qj} in the bulk action.
 For superconformal field theories,
the trace and R-symmetry holographic anomalies are obtained from five-dimensional minimal 
gauged supergravity. The latter arises as a consistent truncation of \emph{e.g.}\ type IIB supergravity on any five-dimensional Sasaki--Einstein manifold~\cite{BuchelLiu}.
Its bosonic action reads 
\bea
S & = & \frac{1}{16\pi G_5}\int \left[ \diff ^5 x \sqrt{- \hat g}\left(\hat R + \frac{12}{\ell^2} - \frac{\ell^2}{3} \hat F_{\mu\nu}\hat F^{\mu\nu} \right)  - 
\frac{8\ell^3 }{27}\hat A\wedge \hat F \wedge \hat F\right]~,
\label{sugrainrigid}
\eea
where $\mu,\nu$ are five-dimensional indices, $\hat R$ denotes the Ricci scalar of a five-dimensional metric $\hat g_{\mu\nu}$, $\hat A$ is the graviphoton gauge field, with field-strength $\hat F =\mathrm{d} \hat A$, and $\ell$ is the radius of the AdS solution. We have decorated all 
five-dimensional quantities with a hat, to distinguish them from the corresponding  four-dimensional objects, that arise as their 
boundary values. The reason for choosing this unconventional normalisation of the gauge field is that its boundary value can be identified with the background gauge 
field $A_m$ that we considered in the previous
 sections.\footnote{To compare with a more standard normalisation in the literature, 
one should set  $ \hat A = \frac{\sqrt{3}}{\ell} A^\mathrm{s}$ (where ``s'' stands for ``standard''). Using this, the  action of minimal gauged supergravity reads
\label{normalsugra} 
\be
S  =  \frac{1}{16\pi G_5}\int \left[ \diff^5 x \sqrt{- \hat  g}\left( \hat R  + \frac{12}{\ell^2} - F^\mathrm{s}_{\mu\nu} F^{\mathrm{s}\, \mu\nu} \right)  - \frac{8}{3\sqrt{3}}  A^\mathrm{s} \wedge  F^\mathrm{s} \wedge F^\mathrm{s}\right]~.
\label{sugragg}
\ee} 
When  evaluated in an asymptotically locally AdS solution the action (\ref{sugrainrigid}) contains the following logarithmically
divergent term
\bea
I_\mathrm{log} &  = &  \frac{\ell^3}{128 \pi G_5}\int \diff ^4 x \sqrt{- g}\left( R_{mn}R^{mn}  - \frac{1}{3} R^2  - \frac{4}{3} F_{mn}  F^{mn} \right)  ~,
\label{pippo}
\eea
where $R_{mn}$ denotes the four-dimensional Ricci tensor of the boundary metric $g_{mn}$, $R$ its Ricci scalar, and $F_{mn}$ the 
field-strength of the boundary value of the field graviphoton, denoted $A_m$. 
Notice that the Chern--Simons term in (\ref{sugrainrigid}) is not
divergent  \cite{Taylor:2000xw}. 
This result was first obtained in~\cite{Taylor:2000xw},\footnote{The normalisation of the gauge field in \cite{Taylor:2000xw} corresponds to $\frac{\ell}{\sqrt{3}} \hat A = \frac{1}{2} A_\mathrm{Taylor}$.} 
 followed by \cite{Kalkkinen:2000uk,Bianchi:2001kw,Martelli:2002sp}.

We can then use the identity
\be
C_{mnpq}C^{mnpq}  - \euler \ =\  2  R_{mn}R^{mn}  - \frac{2}{3} R^2  ~
\ee
to rewrite the divergent action (\ref{pippo}) as 
\bea
I_\mathrm{log} &  = &  \frac{\ell^3}{256 \pi G_5}\int \diff^4 x \sqrt{- g}\left( C_{mnpq}C^{mnpq}  - \frac{8}{3}  F_{mn}  F^{mn}  - \euler \right)  ~.
\label{pluto}
\eea
This  is  proportional to the bosonic action of conformal supergravity \cite{FradkinTseytlin}, up to the topological term
\cite{Liu:1998bu,Balasubramanian:2000pq}.
 Moreover, the integrand of the logarithmically divergent part of the action is 
proportional to the trace anomaly \cite{Henningson:1998gx}. 
More precisely, in the radial variable used in \cite{Henningson:1998gx} the proportionality factor is $-2$,  hence we have
\bea
 T_m^m |_\mathrm{hol} & = & \frac{\ell^3}{128 \pi G_5}\left( C_{mnpq}C^{mnpq}   -\frac{8}{3}  F_{mn}  F^{mn}  - \euler   \right)~.
\eea
Comparing this with (\ref{correcttrace}) yields the well-known large $N$ result~\cite{Henningson:1998gx}
\bea
a \ = \ c & = & \frac{\ell^3\pi }{8G_5}~.
\label{hologac}
\eea
For supersymmetric backgrounds, the relations (\ref{Csq=FsqLorentzian}) imply that the divergent part of the on-shell action is purely topological, namely
\bea
I_\mathrm{log}^\mathrm{susy} &  = & - \frac{\ell^3}{256 \pi G_5}\int \diff^4 x \sqrt{- g} \,  \euler   ~.
\label{pluto}
\eea
Note that this explains the observation 
made in \cite{Taylor:2000xw}, that in the magnetic string solution of \cite{ChamseddineSabra,KlemmSabra} the on-shell action does not have logarithmic divergences. 
For completeness, let us recall how the  well-known formula for the central charge $a$ is obtained from type IIB supergravity on AdS$_5 \times M_5$, where $M_5$ is a Sasaki--Einstein manifold. 
Using $\frac{1}{G_5} 
= \frac{\ell^5\mathrm{vol'}(M_5)}{G_{10}}$, where ${\rm vol'}(M_5)$ is the dimension-less volume of the manifold $M_5$, together with the expression for the AdS$_5$ radius in terms of  $N$ units of five-form flux,
\be\label{MAGOO1}
\ell^4 =  \frac{\sqrt{2G_{10}}\,\pi N}{{\rm vol'}(M_5)}\,,
\ee
eq.~(\ref{hologac}) can be written as
\be
a\ =\ c\ = \frac{\pi^3 N^2 }{4{\rm vol'}(M_5)} \,.
\ee

Let us now discuss  the holographic R-symmetry anomaly.  A  calculation of the holographic chiral anomaly induced by a bulk Chern--Simons term for an Abelian gauge field 
 has appeared in \cite{Landsteiner:2011iq}, and it is straightforward to  extract the R-symmetry anomaly from the results of this reference. 
The authors  of \cite{Landsteiner:2011iq} consider a five-dimensional Einstein--Maxwell model, whose action, in the normalisations of (\ref{sugragg}), contains a Chern--Simons term
\bea
S_\kappa & = & \frac{\kappa }{2\pi G_5}\int \diff ^5 x \sqrt{-  g}\,
\epsilon^{\mu\nu\rho\sigma\lambda} 
A_\mu^\mathrm{s} F_{\nu\rho}^\mathrm{s}  F_{\sigma\lambda}^\mathrm{s} ~.
\label{sugraspain}
\eea
This leads to  the following expression for  holographic chiral anomaly \cite{Witten:1998qj}:
\bea
  \nabla^m J^\mathrm{s}_m |_\mathrm{hol} & = & - \frac{\kappa}{6\pi G_5} \epsilon^{mnpq} F^\mathrm{s}_{mn} F^\mathrm{s}_{pq}~.
\label{chiralcane}
\eea
To obtain the R-symmetry anomaly we need to use  the value  $\kappa = -\frac{1}{4\sqrt{3}}$ of the Chern--Simons coupling fixed by (\ref{sugragg}),  
and to convert to the normalisation of the gauge field adopted in (\ref{sugrainrigid}).
Namely, we need to  rescale the field $A^\mathrm{s}_m=\frac{\ell}{\sqrt{3}}A_m$, and of course also the current $J^\mathrm{s}_m$ as follows from
\bea
J^s_m = \frac{\delta S}{\delta A^s_m} \ =  \ \frac{\sqrt{3}}{\ell} J_m~.
\eea
Finally, using (\ref{hologac}) we obtain
\bea
 \nabla_m J^m |_\mathrm{hol} & = & \frac{2a}{27\pi^2} F \widetilde F~,
\eea
which agrees exactly with (\ref{correctchiral}) upon setting $a=c$.

\section{Discussion}
\label{discutiamo}

In this paper, we have elaborated on properties of four-dimensional field theories in curved backgrounds preserving rigid supersymmetry. 
In particular, we have described a method for determining 
the field-strength of the background gauge field, using the integrability of the (charged) conformal Killing spinor equation. Noting that backgrounds 
solving the old or new minimal rigid supersymmetry equation must satisfy the CKS equation, our results are valid for all the supersymmetric backgrounds considered so far 
in the literature, both in Lorentzian and Euclidean signature. Using our characterisation of the gauge field, we have shown that on any supersymmetric background there exist precise relations
between various curvature invariants. In Lorentzian signature, these relations imply that the terms multiplying the central charge $c$ in the superconformal anomalies
\emph{vanish}. In Euclidean signature, the existence of a chiral solution to the CKS equation is a slightly weaker condition, and it implies that the $c$-term in the trace anomaly becomes topological, rather than zero. 
However, if there exist two solutions with opposite R-charges, then the $c$-anomalies vanish, as in Lorentzian signature; moreover, all the integrated anomalies vanish, including the terms involving the central charge $a$. A converse version of this 
result might be true, so that perhaps if in a compact complex manifold 
the signature and Euler characteristic vanish, then a second CKS with opposite chirality exists. It would be interesting to see whether such statement holds.

One of the motivations for our work was to study further  the relationship of rigid supersymmetry with local supersymmetry, through holography \cite{KTZ,Cassani:2012ri}. For example, we have shown that, when 
evaluated on supersymmetric asymptotically locally AdS solutions, the logarithmically divergent part of the on-shell action of five-dimensional minimal gauged supergravity is a topological density. In particular, this explains why 
 the on-shell action does not contain the logarithmic divergence in all the (few) supersymmetric solutions in the literature (which also turn out to have $\euler=0$). We have also 
 shown that the coefficient of the
 holographic R-symmetry anomaly  agrees nicely with that of the field theory, filling a small gap in the literature.  Armed with the results of
 \cite{KTZ,Cassani:2012ri}, together with those presented in this paper, we would like to address the problem of constructing non-trivial examples of five-dimensional supergravity solutions, dual to superconformal 
 field theories on curved backgrounds.\footnote{These would be analogous to the four-dimensional gravity duals in  
\cite{Martelli:2011fu,Martelli:2011fw,Martelli:2012sz,Martelli:2013aqa}.} There exist well-known topological obstructions for the existence of a five-dimensional  ``filling'' of four-dimensional manifolds. Specifically, an oriented four-manifold bounds an oriented five-manifold if and only if its signature vanishes.
 As we noted, supersymmetric backgrounds with two supercharges of opposite R-charge automatically satisfy this requirement, and 
 therefore are natural candidates for admitting (smooth) gravity duals. Such five-dimensional gravity-duals will be studied in a separate paper~\cite{wip}.

 It will be interesting  to explore the possibility that the simplification of the anomalies could be used to device new methods for computing the central charges $a$ and $c$
 \cite{Shapere:2008zf}. Furthermore, it is natural to extend our approach to other dimensions, in particular this could lead to analogous simplifications of the superconformal anomalies in six-dimensional SCFTs. One could also consider currents of non-R-symmetries coupled to additional background gauge fields, and study how supersymmetry of the background affects their chiral anomalies as well as the new terms arising in the trace and R-current anomalies.

\subsection*{Acknowledgments}
We would like to thank  M.~Bianchi, D.~Freedman, K.~Intriligator, D.~Panov, K.~Skenderis, J.~Sparks,  A.~Tomasiello,   A.~Tseytlin,   B.~Wecht and L.~Wulff 
for useful comments and correspondence, as well as C.~Closset and Z.~Komargodski for stimulating discussions. 
D.C. is supported by an STFC grant ST/J002798/1. D.M. is supported by an ERC Starting Grant -- Grant Agreement N. 304806 --  Gauge-Gravity,
and also acknowledges partial support from an EPSRC Advanced Fellowship EP/D07150X/3 and from the STFC grant ST/J002798/1.

\appendix

\section{Lorentzian vs Euclidean conventions}\label{ConventionsApp}

In Lorentzian signature $(-+++)$, our orientation is fixed by $\epsilon_{0123}= -\epsilon^{0123} = 1$ and ${\rm vol}_4 = e^0\wedge e^1\wedge e^2\wedge e^3$. 
In Euclidean signature $(++++)$, we take $\epsilon_{1234} = 1$ and ${\rm vol}_4= e^1\wedge e^2\wedge e^3\wedge e^4$. 
In both signatures, the charge conjugate of a spinor $\zeta$ is defined by $\zeta^c = B \zeta^*$, where $B$ is the intertwiner such that $B B^\dagger =1$, $BB^* = -1$ and $\gamma_m^* = (-)^\tau B^{-1}\gamma_m B$, with $\tau=0$ in Euclidean signature and $\tau=1$ in Lorentzian signature. In Lorentzian signature, $\gamma_5 = \ii \gamma^0\gamma^1\gamma^2\gamma^3$, while in Euclidean signature $\gamma_5= \gamma^1\gamma^2\gamma^3\gamma^4$. Note that the charge conjugate of a Lorentzian chiral spinor has opposite chirality, while the charge conjugate of an Euclidean chiral spinor has the same chirality. In Lorentzian signature we also define the Dirac conjugate $\overline{\zeta} =\zeta^\dagger \gamma^0$.

To pass from Lorentzian to Euclidean signature, we set $x^0 = ix^4$ and therefore $\gamma^0 = \ii \gamma^4$.
Since in Lorentzian signature $**=-1$ on two-forms, (anti)-self-dual forms $\phi^{\pm}$ are necessarily complex. In our conventions, $\ii * \phi^{\pm} = \pm \phi^{\pm}$. Note that the complex conjugate of a Lorentzian chiral form has the opposite chirality.
In Euclidean signature, $**=+1$ and we define (anti)-self-dual forms by $* \phi^{\pm} = \pm \phi^{\pm}$. Euclidean chiral forms can be complex or real, and complex conjugation does not flip the chirality.

\section{Derivation of the integrability condition}\label{DerivationIntegrability}

In this appendix, we derive the integrability condition~\eqref{IntegrabilityChargedCKS} of the CKS equation~\eqref{ChargedCKS}. 

Taking the commutator of two covariant derivatives on $\zeta_+$ and using~\eqref{ChargedCKS} we obtain
\be\label{integrabilityStep0}
\frac 14 R_{mnpq} \gamma^{pq} \zeta_+ - \ii F_{mn} \zeta_+\ = \  (\gamma_n \nabla_m^A  - \gamma_m \nabla_n^A)\eta_-\,,
\ee
where we set 
\be
\eta_- = \frac 14 \, \nabla^A \!\cdot\zeta_+\,,
\ee and $\nabla^A_m \eta_- = (\nabla_m - iA_m)\eta_-$.
Contracting with $\gamma^n$ we get
\be\label{integrabilityStep1}
-\frac 12 R_{mn}\gamma^n \zeta_+ - \ii F_{mn} \gamma^n\zeta_+\ = \  2\,\nabla^A_m \eta_- + \gamma_m \,\nabla^A \!\cdot\eta_-\,.
\ee
Contracting this with $\gamma^m$ we determine 
\be
\,\nabla^A\!\cdot\eta_- \ = \ -\frac{1}{12}R\, \zeta_+ -\frac \ii 6 F_{mn}\gamma^{mn} \zeta_+\,,
\ee
which substituted back into~\eqref{integrabilityStep1} gives an equation for $\nabla^A_m \eta_-$:
\be\label{eqforeta}
\nabla^A_m \eta_- \ = \ \Big(-\frac 12 S_{mp} - \frac{\ii}{3} F_{mp} \Big)\gamma^p\zeta_+ + \frac{\ii}{12} \gamma_{mpq} F^{pq}\zeta_+\,,
\ee
where $S_{mn} = \frac 12 \left( R_{mn} - \frac 16 R\, g_{mn} \right)$ is the Schouten tensor. This can be used to eliminate $\eta_-$ from the original equation~\eqref{integrabilityStep0}. We observe that the Riemann tensor combines with its contractions in the Schouten tensor to give the Weyl tensor, defined in~\eqref{defWeyl}. After some gamma-matrix algebra, we arrive at~\eqref{IntegrabilityChargedCKS}.
Note that contracting~\eqref{IntegrabilityChargedCKS} with $\gamma^n$ the equation trivializes, as one may expect observing that the same is true for the CKS equation.

It would be interesting to clarify whether eq.~\eqref{eqforeta} for $\eta_-$ implies additional constraints on the background, which are known to be encoded in the CKS equation~\cite{KTZ,Dumitrescu:2012ha,Cassani:2012ri}. For instance, one can consider the integrability condition of~\eqref{eqforeta}, which yields (in Lorentzian signature for definiteness)
\be
\Big(\frac 14 C_{mnpq} - \ii \, g_{p[m} F_{n]q} \Big)\gamma^{pq} \eta_- =  \Big( - \nabla_{[m} S_{n]p} -\frac{2\ii }{3} \nabla_{[m} F_{n]p} + \frac 13 \nabla_{[m} \widetilde F_{n]p} \Big) \gamma^p \zeta_+ + \frac{\ii}{3}\Big( 5 F_{mn} + \ii \widetilde F_{mn} \Big) \eta_-.
\ee
Contracting with $\gamma^n$ and using the twice-contracted Bianchi identity of the Riemann tensor (written in the form $\nabla_{[n}S_{m]}{}^n = 0$) we obtain
\be
\frac{4\ii}{3}F^-_{mn}\gamma^n \eta_- + \Big(\! -\frac \ii 3 \nabla^n F_{nm} + \frac 16 \nabla^n\widetilde F_{nm} \Big)\zeta_+ + \Big(\frac 12 \nabla_n S_{mp} - \frac{\ii}{2} \nabla_{[m}F_{np]} -\frac{1}{6} \nabla_n \widetilde F_{mp} \Big) \gamma^{np}\zeta_+ = 0.
\ee
Contracting again with $\gamma^m$, we eventually arrive at
\be
(\diff F)_{mnp} \gamma^{mnp} \zeta_+ \ = \ 0\,,
\ee
which is automatic if the Bianchi identity for $F$ is satisfied.

\section{Review of anomaly formulae}
\label{reviewanomalies}

In this appendix, we briefly review how the superconformal anomaly formulae are obtained from a superspace computation~\cite{Anselmi:1997am} and give some arguments in support of the corrected expressions~\eqref{correcttrace}, \eqref{correctchiral}. In the limit $a=c$, the validity of these expressions is also confirmed by large $N$ holography as discussed in section~\ref{holosection}.
An introduction to the relevant superspace formalism can be found in~\cite{WessBagger,Superspace,BuchbinderKuzenkoBook}. 

In superspace, the stress-energy tensor $T_{mn}$ and the R-current $J_m$ are components of the Ferrara--Zumino supercurrent  $\mathcal J_{\alpha\dot\alpha}\,$, which is a real vector superfield satisfying the conservation equation
\be
\overline D^{\dot \alpha} \mathcal J_{\alpha\dot\alpha} \ = \ D_\alpha \mathcal T\,.
\ee
The chiral superfield $ \mathcal T$ is called the supertrace, and its $\theta\theta$ component contains both the trace of the stress-energy tensor and the divergence of the R-current:
\be
-\frac 14 (D^\alpha D_\alpha  \mathcal T)|_{\theta = 0} \ =\   \frac 23\, T_m^m + \ii \, \nabla_mJ^m \,,
\ee
where we use the conventions of~\cite{WessBagger}. More complete expressions for these superfields~can be found e.g.\ in~\cite[sect.$\:$2.1]{KomargodskiSeiberg}.
The supertrace vanishes for a superconformal field theory on flat space, but acquires anomalous contributions when the theory is put on a non-trivial background.\footnote{If the classical theory is not conformally invariant then in addition to the quantum contribution~\eqref{SupertraceAnomaly} the supertrace also includes a classical term.}
 One has
(see~\cite{BonoraPastiTonin} for the general form and~\cite{McArthur:1983fk,BuchbinderKuzenkoAnomalies} for an explicit evaluation)
\be\label{SupertraceAnomaly}
 \mathcal T \ = \ \frac{1}{24\pi^2}\left( c\, W^2 - a \, \Xi_c \right),
\ee
where $W^2 = W^{\alpha\beta\gamma}W_{\alpha\beta\gamma}$ is the square of the super-Weyl tensor, while $\Xi_c$ is the chirally projected super-Euler density. Extracting the $\theta\theta$ component gives the relevant anomaly formulae,
\be\label{GeneralExprSConfAnomaly}
\frac 23\, T_m^m + \ii \, \nabla_mJ^m \ =\  \frac{1}{24\pi^2}\left( -\frac c4 \, D^\alpha D_\alpha W^2|_{\theta = 0} + \frac a4 \, D^\alpha D_\alpha\Xi_c|_{\theta = 0} \right).
\ee
The authors of~\cite{Anselmi:1997am} evaluate this expression in their appendix~A, obtaining their equations (4.3) and (4.7). In the following we motivate why that result should be amended, by putting together some expressions appeared elsewhere in the literature. The super-Weyl invariant $D^\alpha D_\alpha W^2$ has been independently evaluated in~\cite{Lu:2011mw}, and is given by eq.~(B.23) therein.
 In our notation, its bosonic part reads\footnote{In the conventions of~\cite{Lu:2011mw} (LPSW), $G_{ab}=2D_{[a}G_{b]}$, with $G_a| = \frac 16 A_a^{\rm LPSW}$. Moreover, their vector field $A^{\rm LPSW}$ is related to our $A$ by $A^{\rm LPSW} = -2 A$. To see this, compare the gravitino variation in their eq.~(2.9) with our old minimal equation~\eqref{OldMinimalEq}, note that $A^{\rm LPSW}$ is the same as $b$, and recall the discussion below our~\eqref{OldMinimalEq}.
}
\be
-\frac 14 (D^\alpha D_\alpha W^2)|_{\theta = 0} \ =\  C_{mnpq}C^{mnpq} - \frac 83 F_{mn} F^{mn} + \ii \,R_{mnpq}\widetilde R^{mnpq} - \frac 83\ii \, F_{mn}\widetilde {F}{}^{mn}.
\ee
Note that the expression on the right hand side is exactly the one that is set to zero by the integrability of the Lorentzian CKS equation (indeed $R_{mnpq}\widetilde R^{mnpq}= C_{mnpq}\widetilde C^{mnpq}$). Hence we have found that, at least in Lorentzian signature, the CKS equation implies vanishing of the bosonic part of the super-Weyl invariant. Plugging the expression above into~\eqref{GeneralExprSConfAnomaly}, we arrive at
\be\label{AnomalyFromLPSW}
\frac 23 T_m^m + \ii \nabla_m J^m = \frac{c}{24\pi^2}\! \left( C_{mnpq}C^{mnpq} - \frac 83 F_{mn} F^{mn} + \ii\, R_{mnpq}\widetilde R^{mnpq} - \frac 83\ii \, F_{mn}\widetilde {F}{}^{mn}\right)  + a\textrm{-terms}.
\ee
Separating the real and imaginary parts, we obtain precisely the $c$-terms in our eqs.~\eqref{correcttrace}, \eqref{correctchiral}. Note the symmetry between the relative coefficients in the real and imaginary parts. These coefficients also agree with the independent superspace computation done in~\cite{SchwimmerTheisen}. Comparing with eqs.~(4.3) and (4.7) of~\cite{Anselmi:1997am}, if we assume that our field-strength $F$ is the same as their $V$, then we find a sign mismatch between the $F^2$ and $V^2$ terms as well as a mismatch by a factor of 1/3 between the $F\widetilde F$ and $V\widetilde V$ terms.

Finally, we compare the expression for the R-current anomaly with the one given in~\cite{IntriligatorWecht} (IW), which was not obtained from superspace. Specializing eq.~(2.8) in~\cite{IntriligatorWecht} to the R-current, we have
\be\label{RanomalyIW}
\nabla_m J^m_{\rm IW} \ = \ \frac{1}{{384 \pi^2}}\left(k_{R}\, R_{mnpq} \widetilde R^{mnpq} + 8k_{RRR}\, F^{\rm IW}_{mn}\,\widetilde F^{{\rm IW}\,mn} \right).
\ee
The anomaly coefficients $k_R$ and $k_{RRR}$ can be eliminated in favour of the SCFT central charges $a$ and $c$ using the relations
$a = \frac{3}{32}(3 k_{RRR} - k_R)$ and $c = \frac{1}{32}(9 k_{RRR} - 5 k_R)$ found in~\cite{Anselmi:1997am}. In this way one obtains
\be\label{JanomalyIW}
\nabla_m J^m_{\rm IW} \ = \  \frac{a-c}{24\pi^2}\, R_{mnpq} \widetilde R^{mnpq} + \frac{5a-3c}{27\pi^2}\, F^{\rm IW}_{mn}\,\widetilde F^{{\rm IW}\,mn},
\ee
which agrees with our eq.~\eqref{correctchiral} provided one redefines $J_m = -J_m^{\rm IW}$ and $F_{mn} = \ii\, F^{\rm IW}_{mn}\,$.\footnote{We thank Ken Intriligator and Brian Wecht for communications on consistency of~\eqref{RanomalyIW} with~(\ref{correctchiral}).}


\end{document}